\documentclass[12pt]{article}
\usepackage{geometry}                % See geometry.pdf to learn the layout options. There are lots.
\usepackage{graphicx}
\usepackage{amssymb}
\usepackage{amsmath}
\usepackage{chemarrow}
\numberwithin{equation}{section}

\def\C{\mathbb{C}}
\def\P{\mathbb{P}}
\def\FFh{\hat {\cal F}}

\def\id{{1 \kern-.28em {\rm l}}}

\def\K3{{\bf K3}}
\def\journal#1&#2(#3){\unskip, \sl #1\ \bf #2 \rm(19#3) }
\def\andjournal#1&#2(#3){\sl #1~\bf #2 \rm (19#3) }

\def\bar{\overline}
\def\hat{\widehat}
\def\ie{{\it i.e.}}
\def\eg{{\it e.g.}}

\def\tilde{\widetilde}

\def\half{\frac12}

\def\d{\partial}

\def\inbar{\,\vrule height1.5ex width.4pt depth0pt}
\def\IC{\relax\hbox{$\inbar\kern-.3em{\rm C}$}}
\def\IR{\relax{\rm I\kern-.18em R}}
\def\IP{\relax{\rm I\kern-.18em P}}

\catcode`\@=11
\def\slash#1{\mathord{\mathpalette\c@ncel{#1}}}
\def\BB{{\cal B}}
\def\BBh{\hat {\cal B}}

\def\FF{{\cal F}}

\def\HH{{\cal H}}

\def\KK{{\cal K}}

\def\OO{{\cal O}}

\def\UU{{\cal U}}

\def\WW{{\cal W}}

\def\vareps{\varepsilon}

\def\underrel#1\over#2{\mathrel{\mathop{\kern\z@#1}\limits_{#2}}}

\catcode`\@=12

\def\det{{\rm det}}
\def\tr{{\rm tr}}

\def \sinh{{\rm sinh}}
\def \cosh{{\rm cosh}}
\def \sgn{{\rm sgn}}
\def\det{{\rm det}}

\def\zbar{{\bar z}}

\def\Zbar{{\bar Z}}

\def\lambdabar{\bar{\lambda}}
\def\etabar{\bar{\eta}}

\def\alphabar{{\bar\alpha}}
\def\betabar{{\bar\beta}}
\def\gammabar{{\bar\gamma}}

\def\abar{{\bar a}}
\def\bbar{{\bar b}}
\def\cbar{{\bar c}}

\def\hbar{{\bar h}}

\def\vbar{{\bar v}}
\def\wbar{{\overline w}}

\def\del{\partial}
\def\delbar{{\bar \del}}
\def\hlambda{\hat{\lambda}}
\def\hlambdabar{\overline{\hat{\lambda}}}
\def\hgamma{\hat{\gamma}}
\def\hgammabar{\overline{\hat{\gamma}}}
\def\Xbar{\overline{X}}
\def\varepsbar{\overline{\varepsilon}}
\begin{document}

\begin{titlepage}

\begin{flushright}
DAMTP-2011-50 \\
RUNHETC-2011-15
\end{flushright}
\vspace*{3mm}
\begin{center}
{\bf\Large T-dualising the Deformed and Resolved Conifold}
\vspace*{7mm}

{Jock~McOrist$^1$ and Andrew~B.~Royston$^2$}
\vspace*{7mm}

{\em ${}^1$DAMTP, Centre for Mathematical Sciences \\ Wilberforce Road, Cambridge, CB3 OWA, UK}

\vspace{3mm}

{\em ${}^{2}$NHETC and Department of Physics and Astronomy \\ Rutgers University, Piscataway, NJ 08855, USA}
\end{center}

\vspace*{20mm}

\begin{abstract}
In a previous paper we used T-duality to construct a new type of $1/4$-BPS solution describing a pair of NS5-branes intersecting in $1+3$ dimensions and localised in all other directions except for a single transverse circle. This led to an explicit solution to a sourced Monge--Ampere equation, of which there are few known  examples. In this paper we refine this formalism and apply it to two important generalisations: the resolved and deformed conifolds. In doing so we construct two new solutions describing, respectively, a pair of NS5-branes separated in a transverse direction and a pair of NS5-branes with smooth `diamond' profile.  We show how the parameter of the resolved conifold (size of the $S^2$) maps to a transverse separation of the NS5-branes, while the modulus of the deformed conifold (size of the $S^3$) maps to the deformation parameter of the diamond web.

\end{abstract}
\end{titlepage}

\tableofcontents

%%%%%%%%%%%%%%%%%%%%%%
%%%%%%%%%%%%%%%%%%%%%%
\section{Introduction} \label{Introduction}
%%%%%%%%%%%%%%%%%%%%%%
%%%%%%%%%%%%%%%%%%%%%%
Solitonic brane solutions of type II supergravity have played an important role in various areas of string theory. A prominent example is understanding strongly coupled field theories via the application of gauge--gravity duality. Another is to use intersecting brane configurations to UV complete effective field theories, so that one can study novel field-theoretic phenomena at the cut-off scale. To that end, it is of interest to study the dynamics of D-brane probes in non-trivial brane backgrounds.  Indeed, the original motivation for this work came from the study of how the gravitational effects of a single NS5-brane influence the dynamics of a D4-brane probe \cite{Giveon:2009ur,Kutasov:2009kb}. This gravitational background is well-known and is given by the CHS solution \cite{Callan:1991at}.  This resulted in innovative mechanisms for realising, say, metastable supersymmetry breaking in the $1+3$ dimensional effective field theory. However, more interesting phenomena may arise when the gravitational interactions of multiple NS5-branes intersecting non-trivially are included. Unfortunately, there are few known examples of intersecting localised brane solutions that preserve at most 1/4-BPS supersymmetry\footnote{See \cite{McOrist:2011in} for references and a brief review.}. Our main goal  then becomes to construct such examples.

This paper is a continuation of \cite{McOrist:2011in} in which a new example of a 1/4-BPS solution of type II supergravity was constructed. This solution corresponds to a pair of intersecting NS5-branes with common worldvolume $\mathbb{R}^{1,3}$, localised in all directions except for a single mutually transverse $S^1$. Such membrane webs are hard to come by as they essentially amount to solving a nonlinear PDE of Monge--Ampere type with source terms. In \cite{McOrist:2011in} we used T-duality and a Legendre transformation to determine such a solution. That is, when dualised along a certain $U(1)$ isometry intersecting NS5-branes turn into the singular conifold, a non-compact Calabi--Yau manifold~\cite{Bershadsky:1995sp}. As the metric for the conifold is known, one is able to construct the NS5-brane metric, $H_3$-field and dilaton explicitly via the Buscher rules~\cite{Buscher:1987sk}. However, there are subtleties in doing this. For example, there are a number of different $U(1)$ isometries along which one may dualise, and care must be taken in choosing the correct one. Furthermore, after applying the Buscher rules, how does one prove the resulting configuration correctly describes a pair of intersecting NS5-branes? This is answered by an S-duality of the work of \cite{Lunin:2008tf}, in which the form that metrics for intersecting NS5-branes must take is determined. The background is completely fixed up to a single function $\KK$, which obeys a Monge--Ampere equation with source terms determined by the brane web. The burden undertaken in \cite{McOrist:2011in} is to show that after a Legendre transformation, the background is of the appropriate form, and a function $\KK$ is constructed that satisfies the Monge--Ampere equation with the correct singularity structure for a pair of intersecting NS5-branes.

In this paper we continue this line of work by applying the techniques developed in \cite{McOrist:2011in} to Calabi--Yau geometries in which the conifold singularity is resolved. One way to do this is via a small resolution: the conifold singularity is blown-up by a $\P^1$ while preserving the Calabi--Yau condition, giving rise to the resolved conifold whose Ricci-flat metric is described in \cite{Candelas:1989js,PandoZayas:2000sq}. Another approach is to perform a complex structure deformation resulting in the deformed conifold whose metric is written down in  \cite{Candelas:1989js,Minasian:1999tt}. We apply the formalism developed in \cite{McOrist:2011in}, further refined here, to construct the dual NS5-brane solutions. The resolved conifold T-dualises into NS5-branes separated along a mutually orthogonal direction. General arguments suggest that the separation of the branes, denoted $\Delta y$, corresponds to the volume of the $\P^1$ on the resolved conifold side and we determine the precise map here. The deformed conifold dualises into a pair of intersecting NS5-branes with a non-trivial profile, known as the diamond web~\cite{Aganagic:1999fe}. Although some of these results have been understood in the literature  based on other arguments (e.g. \cite{Bershadsky:1995sp}), the results we present here are the first explicit construction of the relevant metrics of the NS5-branes, which together with the results of \cite{Lunin:2008tf}, constitute a proof of these T-duality relations at the level of type II supergravity.

Many open questions remain.  The conifold geometries admit multiple $U(1)$ isometries along which one could possibly dualise, and it would be interesting to apply the techniques developed both here and in \cite{McOrist:2011in} to determine what they correspond to. Moreover, the solutions constructed here and in \cite{McOrist:2011in} are not asymptotically flat. It would be fascinating to determine the asymptotically flat brane solutions which have the 1/4-BPS NS5-brane solutions presented here as their near horizon limits. These would also imply the existence of a new asymptotically flat Calabi--Yau geometry that contains the conifold as its near horizon limit. Another burning question is to construct a background corresponding to intersecting stacks of multiple NS5-branes. The usual arguments suggest such a background would be T-dual to an orbifold of the conifold \cite{Bershadsky:1995sp,Uranga:1998vf}, though the Ricci-flat metrics of the resolutions of the orbifold, to the best of our knowledge, are not yet known.

Moving beyond supergravity, a natural direction to pursue is to understand the underlying worldsheet description and the quantum corrections of these examples. This would allow, for example, one to study the completely localised brane solution analogous to the phenomenon discussed for a single stack of NS5-branes in \cite{Tong:2002rq}. As there is a non-trivial dilaton profile, it would also be interesting to lift this configuration to M-theory in which it would describe a non-trivial M5-brane profile. Finally, one could study the dynamics of D-brane probes in this background, going beyond the work of say \cite{Giveon:2009ur}.

%%%%%%%%%%%%%%%%%%%%%%%%%%%%%%%%%%%%%%
%%%%%%%%%%%%%%%%%%%%%%%%%%%%%%%%%%%%%%
\section{NS5-brane webs and their geometric duals} \label{Review}
%%%%%%%%%%%%%%%%%%%%%%%%%%%%%%%%%%%%%%
%%%%%%%%%%%%%%%%%%%%%%%%%%%%%%%%%%%%%%

In this section we review the work of \cite{McOrist:2011in} in which we showed how to map the singular conifold to a pair of intersecting NS5-branes, and set up a general framework for understanding the T-duality relation, at the level of supergravity, between NS5-brane webs and their conifold-like, pure metric duals.  We also describe a slight modification of the near-brane analysis given in \cite{Lunin:2008tf} that is necessary to characterise the brane locus in situations where the induced metric on the brane worldvolume is non-trivial.

%%%%%%%%%%%%%%%%%%%%%%%%%%%%%%
\subsection{The geometry of $1/4$-BPS NS5-brane webs}
%%%%%%%%%%%%%%%%%%%%%%%%%%%%%%

 We are interested in a class of ten-dimensional type II supergravity backgrounds containing NS5-branes.  The NS5-branes overlap in $1+3$ dimensions; we use coordinates $x^\mu = (t,{\bf x})$ to parameterise the intersection.  In order to visualise the remaining six directions it is useful to first consider the case of two orthogonally intersecting stacks of branes.  The six directions are split into three pairs: one pair, $(x^4,x^5)$, is tangential to the first stack and transverse to the second; another pair, $(x^8,x^9)$, is transverse to the first and tangential to the second; and the final pair, $(x^6,x^7)$, is orthogonal to both stacks.  We use a complex coordinate system for the relatively transverse directions, $z^a = (z^1,z^2)$, with $z^1 = x^4+i x^5$ and $z^2 = x^8+ix^9$, while the overall transverse directions are denoted by ${\bf y} = (y^i)$.  The class of configurations we consider includes more general profiles---or ``webs''---of branes situated on two-dimensional slices of the four-dimensional space spanned by $(z^1,z^2)$.

The supergravity background generated by an NS5-brane web was determined in \cite{McOrist:2011in} by U-dualising results of \cite{Lunin:2008tf}.  It consists of a non-trivial metric, dilaton, and Neveu--Schwarz three-form flux, all determined in terms of a single function $\KK = \KK({\bf y},z^a,\zbar^\abar)$:
\begin{align}\label{NSfive}
& ds^2  = - dt^2 + d{\bf x}_{3}^2 + 2 \KK_{a \bbar} dz^a d\zbar^\bbar  + e^{-3A} d{\bf y}_{2}^2~, \cr
& e^{\varphi-\varphi_0} = e^{-3A/2}~, \cr
& H_3 = - i \left( \d_a e^{-3A} dz^a - \delbar_{\abar} e^{-3A} d\zbar^\abar \right) \wedge d^2 {\bf y} - i \epsilon_{i}^{\phantom{i}j} \d_j \KK_{a\bbar} dy^i dz^a d\zbar^\bbar~,
\end{align}
with
\begin{align}\label{gAK}
& \KK_{a\bbar} = \d_a \delbar_{\bbar} \KK~, \cr
& \frac{1}{4} e^{-3 A} =  \d_1 \delbar_{\bar 1} \KK \d_2\delbar_{\bar 2} \KK - \d_1\delbar_{\bar 2} \KK \d_2 \delbar_{\bar 1} \KK \equiv \det{(\d \delbar \KK)}~.
\end{align}
Here $H_3 = dB_2$ is the field strength of the NS-NS two-form, and $e^{\phi_0} = g_s$ is the asymptotic string coupling.  This parameterisation follows from a systematic analysis of the BPS constraints applied to a general supergravity ansatz consistent with the bosonic symmetries of the system \cite{Lunin:2008tf}.  In particular, no information concerning the profile of the brane web in the four-dimensional space spanned by $z^a$ has been used thus far.

In order to obtain an equation for $\KK$ alone, one must study the supergravity equations of motion.  It is at this point that precise knowledge of the brane profile is required; the branes provide source terms for the metric, dilaton, and NS-NS flux.  The singularity structure of $\KK({\bf y},z^a,\zbar^\abar)$ is determined by these source terms and encodes the location and number of branes in the system.  A highly non-trivial result of \cite{Lunin:2008tf} is that consistency of the sourced equation of motion for $\KK$ with supersymmetry imposes constraints on the type of brane profiles that are possible.  It was demonstrated that these consistency conditions imply holomorphicity of the brane profile: the source locus must be describable as the solution set of a holomorphic equation in $z^a$: $\lambda(z^1,z^2) = \lambda_0$.

The same result also follows from a supersymmetry analysis of probe branes in flat space, but it is important that the result of \cite{Lunin:2008tf} was derived without using any assumptions or input from the probe brane picture.  In the context of D-brane webs, for example, this demonstrates the equivalence between open string and closed string descriptions of the web.

In Appendix A we revisit the supergravity derivation of holomorphic profiles from consistency of the sourced equation of motion for $\KK$ and improve upon the analysis of \cite{Lunin:2008tf}, where an unnecessary assumption concerning the near-brane behaviour of the warp factor, $e^{-3A}$, was made\footnote{We are indebted to O. Lunin for helpful correspondence on this point.}.  For the case of lower-dimensional brane webs, we reproduce the results of \cite{Lunin:2008tf} and demonstrate the validity of the initial assumption which, in addition to holomorphic profiles, implies that the induced metric on the worldvolume of the brane web is flat.  For the NS5-brane webs of interest in this paper however, we show that the initial assumption concerning the warp factor is incorrect in general, and relaxing it allows for the possibility of a non-trivial induced metric on the brane worldvolume while the profile remains holomorphic.  Due to the technical character of these arguments, they are relegated to the appendix while here, and in the next section, we will present the results of the analysis that will be used in the rest of the paper.

The equation determining $\KK$ can be derived from the supergravity equation of motion for the NS-NS flux.  It is a nonlinear PDE of Monge--Ampere type, with source term:
\begin{equation}\label{sourcedMAL}
\Delta_{\bf y} \KK + 8 \det{(\d \delbar \KK)} = -\frac{Q_0}{2\pi} \delta^{(2)}({\bf y} - {\bf y}_0) \log{|\lambda(z^a) - \lambda_0|^2}~,
\end{equation}
where $\Delta_{\bf y}$ is the (flat-space) Laplacian on $\mathbb{R}^2$.  Here we are considering a single NS5-brane web of charge $Q_0$, located at position ${\bf y} = {\bf y}_0$ and described by holomorphic profile $\lambda(z^a) = \lambda_0$.  A web with NS charge one, corresponding to a single NS5-brane, has $Q_0 = (2\pi \ell_{s})^2$.  Multiple webs at different positions in ${\bf y}$ are mutually BPS and the equation of motion for the general case is obtained by superposing the sources.

Equation \eqref{sourcedMAL}, together with \eqref{NSfive} and \eqref{gAK} describes the $1/4$-BPS supergravity background produced by this web.  We note that \eqref{NSfive}-\eqref{sourcedMAL} is invariant under ${\bf y}$-fibred K\"ahler transformations of the form
\begin{equation}\label{KT}
\KK \to \KK + 2 {\rm Re}(f({\bf y},z^a))~,
\end{equation}
where $f$ is a holomorphic function of $z$ and harmonic in ${\bf y}$: $\Delta_{\bf y} f = 0$.

In order to perform the T-duality described in the next section, it will be necessary to have an explicit expression for the NS-NS two-form potential, $B_2$.  In \cite{McOrist:2011in} we found
\begin{align}\label{B2}
& B_2 = \frac{i}{2} \epsilon_{i}^{\phantom{i} j} \left( \KK_{ja}^{\rm reg} dz^a - \KK_{j\abar}^{\rm reg} d\zbar^{\abar} \right) dy^i~, \qquad \textrm{with} \\  \label{Kreg}
& \KK^{\rm reg} \equiv \KK + k~,
\end{align}
where the function $k({\bf y},z^a,\zbar^{\abar})$ satisfies\footnote{In \cite{McOrist:2011in} we took $k$ to be of the form $k({\bf y},z^a,\zbar^{\abar}) = \tilde{f}({\bf y}) \log{|\lambda - \lambda_0|^2}$, where $\tilde{f}$ satisfies $\Delta_{\bf y} \tilde{f} = \frac{Q_0}{2\pi} \delta^{(2)}({\bf y} - {\bf y}_0)$, but there are more general solutions to \eqref{littlek}.}
\begin{equation}\label{littlek}
\Delta_{\bf y} k = \frac{Q_0}{2\pi} \delta^{(2)}({\bf y} - {\bf y}_0) \log{|\lambda - \lambda_0|^2}~, \qquad \d_a \delbar_{\bbar} k = 0~.
\end{equation}
In \eqref{B2} we are using the shorthand $f_i \equiv \d_{y^i} f$ to denote $y^i$-derivatives.  The correction term $k$ in \eqref{Kreg} is necessary in order that $dB_2 = H_3$ in the plane ${\bf y} = {\bf y}_0$, but away from the sources.

Since NS5-branes are magnetic sources for the NS-NS potential, $dH_3 \neq 0$ on the brane locus $({\bf y},\lambda) = ({\bf y}_0,\lambda_0)$, and $B_2$ cannot be defined there.  This is manifested in the fact that $k$ is only defined away from the brane locus: the two conditions \eqref{littlek} are inconsistent with each other at $({\bf y}, \lambda) = ({\bf y}_0, \lambda_0)$.  Since $k$ is not defined there, neither is $\KK^{\rm reg}$.   $H_3$ can only be trivialised on the complement of the brane locus, and \eqref{B2} gives such a trivialisation.  Finally we note that $B_2$ is shifted by an exact term under the transformations \eqref{KT}.

%%%%%%%%%%%%%%%%%%%%%%%%
\subsection{The pure metric T-dual}
%%%%%%%%%%%%%%%%%%%%%%%%

In \cite{McOrist:2011in} we T-dualised the brane web geometry \eqref{NSfive}-\eqref{sourcedMAL} by taking one of the transverse directions to be a circle, smearing the configuration to create a $U(1)$ isometry, and applying Buscher's T-duality rules for supergravity \cite{Buscher:1987sk,Bergshoeff:1995as}.  Specifically, let $x^7$ have asymptotic periodicity $x^7 \sim x^7 + 2\pi R_7$, and denote the remaining transverse direction $x^6 = y$.  Smear the source terms in \eqref{sourcedMAL}, \eqref{littlek} by replacing $\delta^{(2)}({\bf y}) \to \frac{1}{2\pi R_7} \delta(y)$.  We may assume that $\KK$ is independent of $x^7$, and then the configuration \eqref{NSfive} possesses a $U(1)$ isometry corresponding to translations in $x^7$.

A careful application of Buscher's rules yields a dual supergravity background that is pure metric, with
\begin{align}\label{gcnice}
& d\tilde{s}^2 = -dt^2 + d{\bf x}_{3}^2 + d \tilde{s}_{6}^2~, \qquad \textrm{where} \cr
& d \tilde{s}_{6}^2 = e^{-3A} dy^2 + 2 \KK_{a\bbar} dz^a d\zbar^{\bbar} + e^{3A} \left[ d\tilde{x}^7 - \frac{i}{2} \left( \KK_{ya}^{\rm reg} dz^a - \KK_{y\abar}^{\rm reg} d\zbar^{\abar} \right) \right]^2~.
\end{align}
The dual circle is parameterised by $\tilde{x}^7$ and has asymptotic radius $\tilde{R}_7 = \ell_{s}^2/R_7$.  The warp factor is determined in terms of the potential as before, $e^{-3A} = 4 \det{(\d \delbar \KK)}$, and the equation of motion for $\KK$ is
\begin{equation}\label{sourcedMAS}
\d_{y}^2 \KK + 8 \det{(\d \delbar \KK)} = - \frac{\tilde{Q}_0}{2\pi} \delta(y-y_0) \log{|\lambda(z^a) - \lambda_0|^2}~.
\end{equation}
Here $\tilde{Q}_0 = Q_0/(2\pi R_7)$ is the charge density of the smeared brane web in the original geometry.  In the dual geometry, \eqref{gcnice}, it plays the role of a topological charge.  Note that for a charge one brane web, $\tilde{Q}_0 = 2\pi \ell_{s}^2/R_7 = 2\pi \tilde{R}_7$.  $\KK^{\rm reg}$ is defined in terms of $\KK$ via
\begin{align}\label{littlekS}
& \KK^{\rm reg} = \KK + k~, \qquad \textrm{with} \cr
& \d_{y}^2 k = \frac{\tilde{Q}_0}{2\pi} \delta(y-y_0) \log{|\lambda - \lambda_0|^2}~, \qquad \d_a \delbar_{\bbar} k = 0~.
\end{align}

Let us discuss some global aspects of the geometry described by \eqref{gcnice}.  Consider the transformations \eqref{KT}, where we restrict $f$ to be $x^7$-independent in order to preserve the isometry.  Then under
\begin{equation}\label{KT2}
\KK \to \KK + 2 y {\rm Re}(f_1(z^a)) + 2 {\rm Re}(f_0(z^a))~,
\end{equation}
we have $\frac{i}{2} (\delta \KK_{ya} dz^a - \delta \KK_{y\abar} d\zbar^{\abar}) = -d ({\rm Im} f_1)$.  The metric \eqref{gcnice} remains invariant provided we simultaneously shift ${\tilde x}^7$ according to
\begin{equation}\label{x7shift}
\tilde{x}^7 \to \tilde{x}^7{}' = \tilde{x}^7 - {\rm Im}f_1~.
\end{equation}
Gluing together coordinate patches with such transitions on the overlaps can lead to non-trivial circle fibrations.

Indeed, we know that non-trivial circle fibrations must be present.  The brane web we started with has a conserved charge, $Q_0$, that can be computed via $\int H_3$ over any three-cycle, $\Sigma_3$, enclosing the web.  In the smeared case the charge is $\tilde{Q}_0$ and the three-cycle has the form $\Sigma_3 = S^1 \times \Sigma_2$, where $S^1$ is the $x^7$ circle and $\Sigma_2$ is any two-cycle enclosing the smeared web.  Under T-duality, the conserved NS charge maps to a topological charge.  In the simplest example of a single straight NS5-brane the T-dual geometry is $d\tilde{s}_{6}^2 = \mathbb{C} \times TN$, where $TN$ is a single-centered Taub-NUT space.  The circle fibre corresponds to the Hopf fibre and the topological charge is measured by the Hopf invariant.  This picture can be extended to smooth charge $n$ webs, where one expects the geometry to look locally like a $\mathbb{C}$ fibration over $n$-centered Taub-NUT, or equivalently an $A_{n-1}$ singularity when $n > 1$.

Strictly speaking, the circle fibre in Taub-NUT is only non-trivial when we remove the NUT point where the fibre shrinks to zero.  A related fact is that the $B$-field produced by NS5-branes is only defined on the complement of the brane locus, so the T-duality map is only defined on the complement.  T-duality maps the complement of the brane locus to the complement of the $\mathbb{C}$ fibres over the NUT points.  The original brane web geometry, \eqref{NSfive}, is singular at the locations of sources.  As long as the NS5-branes are not coincident, however, the T-dual geometry can be completed to a regular geometry by adding in the image of the brane locus.  This is well known in the simplest case of Taub-NUT, where the neighbourhood of the NUT point is completely smooth and diffeomorphic to $\IR^4$.  We will see that it follows in the general case of a smooth charge one web from our near-brane analysis of $\KK$ described below.

Finally, let us comment on the supersymmetry of \eqref{gcnice}.  It follows from the work of \cite{Lunin:2008tf} that the brane web geometry \eqref{NSfive} is $1/4$-BPS.  T-duality does not preserve supersymmetry in general, but a sufficient condition for the preservation of supersymmetry is that the Killing spinors of the original solution be independent of the coordinate parameterising the T-duality direction \cite{Bakas:1994ba,Alvarez:1995zr,Alvarez:1995ai}.  Since we have explicitly smeared the original geometry along the T-duality direction, one certainly expects the Killing spinors to satisfy this criteria.  Nonetheless in \cite{McOrist:2011in} we explicitly verified that \eqref{gcnice}, with \eqref{sourcedMAS}, is indeed $1/4$-BPS.  This implies that the six-dimensional space with metric $d\tilde{s}_{6}^2$ must be Calabi--Yau.  In the next section we will show how to map \eqref{gcnice} to a Ricci-flat K\"ahler metric.

%%%%%%%%%%%%%%%%%%%%%%%%%%%%%%%%%%%%%%%%%%
\subsection{Legendre transformations away from the source locus} \label{Legendre}
%%%%%%%%%%%%%%%%%%%%%%%%%%%%%%%%%%%%%%%%%%

Let us begin by analyzing the function $\KK^{\rm reg}$ more closely, which we recall is well defined away from the source locus\footnote{``Source locus'' may refer to the actual brane locus on the brane web side of the T-duality map or the image of the brane locus on the geometric side; here it is the latter.  In either case it is the location of the source terms on the right-hand side of the PDE determining the function $\KK$.} at $(y,\lambda) = (y_0,\lambda_0)$. We have $\d_a \delbar_{\bbar} \KK^{\rm reg} = \d_a \delbar_{\bbar} \KK$, and hence the warp factor can be expressed as $e^{-3A} = 4 \det{(\d \delbar \KK^{\rm reg})}$.  It follows that both the brane web geometry, \eqref{NSfive}, and the T-dual geometry, \eqref{gcnice}, are completely characterised by $\KK^{\rm reg}$ on the complement of the source locus.  Furthermore,
\begin{align}\label{KregMA}
\d_{y}^2 \KK^{\rm reg} + 8 \det{(\d \delbar \KK^{\rm reg})} =&~ \d_{y}^2 \KK + \frac{\tilde{Q}_0}{2\pi} \delta(y-y_0) \log{|\lambda - \lambda_0|^2} +  8 \det{(\d \delbar \KK)} \cr
=&~ 0~, \qquad \qquad \textrm{(away from sources),}
\end{align}
where \eqref{littlekS} and \eqref{sourcedMAS} have been used.  From this we have $\d_{y}^2 \KK^{\rm reg} = \frac{1}{2} e^{-3A}$, and thus we arrive at the following alternative description of the space \eqref{gcnice}:
\begin{align}\label{gcnice2}
& d\tilde{s}_{6}^2 = - \frac{1}{2} \KK_{yy}^{\rm reg} dy^2 + 2 \KK_{a\bbar}^{\rm reg} dz^a d\zbar^{\bbar} - \frac{2}{\KK_{yy}^{\rm reg}} \left[ d\tilde{x}_7 - \frac{i}{2} \left( \KK_{ya}^{\rm reg} dz^a - \KK_{y\abar}^{\rm reg} dz^{\abar} \right) \right]^2~, \\ \label{MAnice}
& \KK_{yy}^{\rm reg} + 8 \det{(\d \delbar \KK^{\rm reg})} = 0~, \qquad \qquad \qquad \textrm{(away from sources).}
\end{align}
In \cite{McOrist:2011in} we showed that a metric of this form is equivalent to a Ricci-flat K\"ahler metric with a $U(1)$ isometry.  Here we review the key points.

Consider a K\"ahler manifold parameterised by coordinates $(X,z^a)$.  Suppose that the manifold possesses a $U(1)$ isometry, which may have fixed points.  We will say more about the nature of the fixed points below; for now we work away from any fixed points and we choose the coordinate system $(X,z^a)$ in a local patch such that the isometry is identified with the phase of $X$.  Let us introduce dimensionless real coordinates $\xi \in \mathbb{R}$, $\phi_X \in [0,2\pi)$, and a real constant $c$ of length dimension one, such that $X = c e^{\xi/2} e^{i\phi_X}$.  We may then assume that the K\"ahler potential, $\FF$, is independent of $\phi_X$ so that
\begin{align}\label{gKahler}
\FF =&~ \FF(\xi, z^a,\zbar^{\abar})~, \cr
ds^2 =&~ 2 \left[ g_{X\Xbar} dX d\Xbar + g_{X\abar} dX d\zbar^{\abar} + g_{a\Xbar} dz^a d\Xbar + g_{a\bbar} dz^a d\zbar^{\bbar} \right]  \cr
=&~ c^2 e^\xi g_{X\Xbar} \left( \frac{1}{2} d\xi^2 + 2 d\phi_{X}^2 \right) + 2 g_{a\bbar} dz^a d\zbar^{\bbar} + \cr
& \qquad \qquad \qquad \qquad \qquad  + \left[ c e^{\xi/2} e^{i\phi_X} g_{X\abar} (d\xi + 2i d\phi_X) d\zbar^{\abar} + c.c. \right] ~,
\end{align}
where $g_{\alpha\betabar} = \d_\alpha \d_{\betabar} \FF$, with $z^\alpha = (X,z^a)$.

Now consider the change of variables
\begin{align}\label{xitoy}
& (\xi, \phi_X, z^a, \zbar^{\abar}) \mapsto (y, \tilde{x}^7, z^a, \zbar^{\abar})~, \qquad \textrm{with} \cr
& \tilde{x}^7 = \pm c \phi_X~, \qquad y= y(\xi,z^a,\zbar^{\abar})~,
\end{align}
where the function $y$ is given by
\begin{equation}\label{Legendre1}
c (y-y_0) = \pm \d_\xi \FF~,
\end{equation}
and define the regularised potential $\KK^{\rm reg}$ as the Legendre transform of $\FF$ with respect to the dual pair $(y,\xi)$:
\begin{equation}\label{Legendre2}
\KK^{\rm reg}(y,z^a,\zbar^{\abar}) = \displaystyle\bigg\{ \FF(\xi,z^a,\zbar^{\abar}) \mp c (y-y_0) \xi \displaystyle\bigg\}_{\textrm{max $\xi$}}~.
\end{equation}
The signs in \eqref{xitoy}-\eqref{Legendre2} are correlated.  In a given coordinate patch they are just a matter of convention, but relative signs between coordinate patches will play an important role in the global analysis discussed below.  It is straightforward to verify that after changing variables according to \eqref{xitoy}, and replacing partial derivatives of $\FF$ with partial derivatives of $\KK^{\rm reg}$ according to \eqref{Legendre1}, \eqref{Legendre2}, the metrics \eqref{gKahler} and \eqref{gcnice2} are equivalent.

The Ricci tensor on a K\"ahler manifold is given by $R_{\alpha\betabar} = \d_\alpha \delbar_{\betabar} \log{g}$, where $g = \det{(g_{\delta\gammabar})}$ is the determinant of the Hermitian metric.  Using \eqref{xitoy}-\eqref{Legendre2}, we find that the determinant can be expressed as
\begin{equation}\label{detgKahler}
g =   - \frac{e^{-\xi}}{\KK_{yy}^{\rm reg}} \det{(\d \delbar \KK^{\rm reg})}~.
\end{equation}
Thus if $\KK^{\rm reg}$ satisfies \eqref{MAnice}, then $g = e^{-\xi}/8 = c^2/(8 |X|^2)$, implying $R_{\alpha\betabar} = 0$.  (Recall that we are explicitly working away from any fixed points of the $U(1)$ isometry, so $X \neq 0$.)  The implication goes the other way as well.  Ricci-flatness for a K\"ahler manifold generically means $g = |F(X,z^a)|^2$ for some holomorphic function $F$.  If the manifold has a $U(1)$ isometry associated with the phase of $X$, then $g = |X|^{2B} |f(z^a)|^2$ for some constant $B$ and holomorphic function $f$.  After a possible rescaling of $X$ and holomorphic change of variables $z^a \to {z'}^{a'}$, this may be brought to the form $g = c^2/8|X|^2$.  Given the identifications \eqref{xitoy}, \eqref{Legendre1}, and \eqref{Legendre2}, equation \eqref{MAnice} then follows.  It is not surprising that a coordinate reparameterisation may be necessary to put Ricci-flatness in the form \eqref{MAnice}, as this equation is not covariant under coordinate transformations.

We have provided a prescription for how to rewrite \eqref{gcnice2}, \eqref{MAnice} as a Ricci-flat K\"ahler metric with $U(1)$ isometry and vice-versa.  The prescription is purely local since we are working in a given coordinate patch, away from any fixed points of the isometry.  There are two elements that are required to extend it globally.  First, we should understand how the transformations \eqref{KT2}, \eqref{x7shift} are realised in the K\"ahler geometry \eqref{gKahler}.  These transformations are expected to play an important role in providing a global definition of the circle fibre parameterised by $\tilde{x}^7$.  Second, we should characterise the behaviour of the geometry in the vicinity of the fixed point locus of the $U(1)$ isometry.  The relationship \eqref{xitoy} between $\phi_X$ and $\tilde{x}^7$ suggests that the fixed point locus should be identified with the source locus of the Monge--Ampere equation, \eqref{sourcedMAS}, where the circle fibre shrinks to zero.  In the remainder of this section we describe how the transformations \eqref{KT2}, \eqref{x7shift} are realised in the K\"ahler geometry, while in the next we study the behaviour of the geometry near the source locus.

Suppose we are on the overlap of two coordinate patches, one parameterised by $(X,z^a)$ and the other by $(X',z^a)$, where $X' = X'(X,z^a)$ has the form $X' = X^{\pm 1} e^{-f_1(z^a)/c}$.  Then $\xi' = \pm \xi - \frac{2}{c} {\rm Re}f_1$, and therefore $\d_\xi \FF = \pm \d_{\xi'} \FF$ since these derivatives are taken with $z^a$ held fixed.  It follows that, by choosing the sign of \eqref{Legendre1} appropriately, the definition of $y$ is unchanged--\ie\ $y = y'$.  Meanwhile, choosing the sign of \eqref{xitoy} appropriately, we have that ${\tilde x}^7{}' = \tilde{x}^7 - {\rm Im}f_1$.  In summary,
\begin{equation}\label{yKTs}
X' = X^{\pm 1} e^{-f_1(z^a)/c} \quad \Rightarrow \quad \begin{array}{l} \KK \to \KK' = \KK + 2 y {\rm Re}f_1 \\ \tilde{x}^7 \to \tilde{x}^7{}' = \tilde{x}^7 - {\rm Im} f_1 \end{array}~,
\end{equation}
where we have also chosen the sign in \eqref{Legendre2} for $\KK'$ to match the choices in \eqref{xitoy}, \eqref{Legendre1} for $\tilde{x}^7{}',y'$.  This reproduces the $y$-dependent part of \eqref{KT2} and \eqref{x7shift}.  The $y$-independent piece of \eqref{KT2} is generated by the $X$-independent K\"ahler transformation $\FF \to \FF + 2{\rm Re}( f_0(z^a))$.

Note that the T-duality map makes it clear that it should be possible to define the coordinates $(y,z^a,\zbar^{\abar})$ globally on the non-compact Calabi--Yau, since they were globally defined in the original brane web system.  Only the $\tilde{x}^7$ circle fibre requires a patchwise definition with non-trivial transition functions (which is directly related to the patchwise structure of the $B$-field in the brane web system).  Thus the Ricci-flat K\"ahler geometry dual to an NS5-brane web discussed above must have the following property: there are patches $\UU_i$ with local coordinates $(X_i, z^a)$, where the $z^a$ are to be identified on each patch, such that the transitions are of the form \eqref{yKTs}.  Furthermore the union of these patches should cover everything except complex-dimension one curves.  These curves are the image of the brane locus under the T-duality map.

A class of such geometries is provided by the generalised conifold and its various deformations, which are of course well known to be T-dual to systems of intersecting NS5-branes \cite{Bershadsky:1995sp}.  The (singular) generalised conifold is defined algebraically as the solution set of the equation $x u = v^m w^n$ in $\mathbb{C}^4$, where $m,n$ are positive integers.  We let $(z^1,z^2) = (v,w)$.  When $x \neq 0$ we can solve the equation for $u$ and take $(x,v,w)$ as the complex coordinate system.  Similarly, when $u \neq 0$, $(u,v,w)$ provide a good coordinate system.  The transition function on the overlap, $x = u^{-1} v^m w^n$ indeed has the form \eqref{yKTs}, and the points missed by these two patches are when $x = u = 0$, implying $v$ or $w = 0$.

%%%%%%%%%%%%%%%%%%%%%%%%%%%%%%%%
\subsection{Behaviour near the source locus} \label{nearbrane}
%%%%%%%%%%%%%%%%%%%%%%%%%%%%%%%%

In the previous section we have shown that $d\tilde{s}_{6}^2$ is a Ricci-flat K\"ahler manifold on the complement of the source locus and described how the brane potential $\KK$ is related to the corresponding K\"ahler potential $\FF$.  Here we will study the behaviour of $\KK$ and its derivatives near the locus and show that the geometry \eqref{gcnice} admits a smooth extension to the locus for charge one webs.  We work in a local patch in which we assume the locus is described by a smooth embedding; our analysis does not apply, for example, at the intersection point of two orthogonal branes.

First define the coordinate $\eta$ in a neighbourhood of the locus such that the holomorphic change of variables $(z^1,z^2) \mapsto (\eta,\lambda)$ has unit determinant.  We zoom in towards the locus by letting
\begin{equation}\label{epsdef}
y - y_0 = \epsilon \hat{y}~, \qquad \lambda - \lambda_0 = \epsilon \hlambda~,
\end{equation}
where $(\hat{y},\eta,\hlambda)$ are order one and $\epsilon$ is small.  Then, as discussed in the appendix, $\KK$ has an expansion of the form
\begin{equation}\label{K1K2}
\KK = \KK_1(\eta,\etabar) + \KK_2(y, \eta,\lambda,\etabar,\lambdabar) + \OO(\epsilon^2)~,
\end{equation}
where $\KK_1$ is finite and regular on the locus, while $\KK_2$ is $\OO(\epsilon)$ but has singular second derivatives.  (Terms of order $\epsilon \log{\epsilon}$ are implicitly included in $\KK_2$.)  Plugging this expansion into \eqref{sourcedMAS} we find, at $\OO(\epsilon^{-1})$,
\begin{equation}\label{Poisson}
\left[ \d_{y}^2 + 8 (\KK_1)_{\eta\etabar} \ \d_\lambda \delbar_{\lambdabar} \right] \KK_2 = - \frac{\tilde{Q}_0}{2\pi} \delta(y-y_0) \log{|\lambda -\lambda_0|^2}~.
\end{equation}
Since this is a PDE in $(y,\lambda,\lambdabar)$ only, $(\KK_1)_{\eta\etabar}$ may be viewed as a constant.  The equation is solved by
\begin{align}\label{K2gensol}
\KK_2 =&~ \frac{\tilde{Q}}{2\pi} \displaystyle\biggl\{ D - |y-y_0| \log{ \left[ \sqrt{2 (\KK_1)_{\eta\etabar} } \left( |y-y_0| + D \right) \right]} \displaystyle\biggr\} + \cr
& \qquad \qquad \qquad + (y-y_0) h_1(\eta,\etabar) + 2 {\rm Re}\left[ (\lambda - \lambda_0) h_2(\eta,\etabar) \right]~, \quad \textrm{with} \cr
D =&~ \sqrt{ (y-y_0)^2 + \frac{|\lambda - \lambda_0|^2}{2 (\KK_1)_{\eta\etabar}} } ~,
\end{align}
and where $h_{1,2}(\eta,\etabar)$ are arbitrary.

Using this solution we can straightforwardly construct the leading behaviour of the metric \eqref{gcnice} in the vicinity of the source locus.  We record here the partial derivatives of $\KK$ for future reference:
\begin{align}\label{Kpartials}
& \KK_{yy} =  - \frac{\tilde{Q}_0}{2\pi} \left[  \delta(y - y_0) \log{|\lambda - \lambda_0|^2} + \frac{1}{D} \right] + \OO(1)~, \cr
& \KK_{y\lambda} = \frac{\tilde{Q}_0}{4\pi \lambda} \left( \frac{y}{D} - \sgn{(y)} \right) + \OO(1)~, \cr
& \KK_{y \eta} = - \frac{\tilde{Q}_0}{4\pi} \left( \frac{y}{D} \d_\eta \log{(\KK_1)_{\eta\etabar}} - 2 \d_\eta h_1 \right) + \OO(\epsilon)~, \cr
& \KK_{\lambda\lambdabar} =  \frac{\tilde{Q}_0}{16\pi (\KK_1)_{\eta\etabar} D } + \OO(1)~,  \cr
& \KK_{\lambda\etabar} = - \frac{\tilde{Q}_0 }{16 \pi (\KK_1)_{\eta\etabar} D} \lambdabar \ \delbar_{\etabar} \log{(\KK_1)_{\eta\etabar}} + \delbar_{\etabar} h_2 + \OO(\epsilon)~, \cr
& \KK_{\eta\etabar} = (\KK_1)_{\eta\etabar} + \frac{\tilde{Q}_0}{2\pi} \displaystyle\bigg\{ - \half \d_\eta \delbar_{\etabar} \log{(\KK_1)_{\eta\etabar}} + \frac{|\lambda|^2 |  \d_\eta \log{(\KK_1)_{\eta\etabar}} |^2}{8 (\KK_1)_{\eta\etabar} D} + \cr
& \qquad \qquad \qquad \qquad   + (y-y_0) (h_1)_{\eta\etabar} + 2 {\rm Re}\left[ (\lambda - \lambda_0)(h_2)_{\eta\etabar} \right] \displaystyle\biggr\} + \OO(\epsilon^2)~,
\end{align}
as well as the warp factor,
\begin{equation}\label{warpnb}
e^{-3A} = 4 \det{(\d \delbar \KK)} = \frac{\tilde{Q}_0}{4\pi D } + \OO(1)~.
\end{equation}
After plugging these into \eqref{gcnice}, we make the non-holomorphic change of variables $(\eta, \lambda, \etabar, \lambdabar) \mapsto (\eta,\gamma,\etabar,\gammabar)$, where $\gamma = \gamma(\lambda,\eta,\etabar)$ is given by
\begin{equation}\label{gammanh}
\gamma = \frac{\lambda - \lambda_0}{\sqrt{2(\KK_1)_{\eta\etabar}} }~.
\end{equation}
We then introduce radial-angular variables in the space transverse to the locus,
\begin{equation}\label{radialangular}
y - y_0 = r \cos{\theta}~, \qquad \gamma = r \sin{\theta} e^{i\phi}~, \qquad \tilde{x}^7 = \tilde{R}_7 \psi~,
\end{equation}
where $\psi$ has periodicity $2\pi$.  In these coordinates small $\epsilon$ corresponds precisely to small $r$.  Writing $\tilde{Q} = 2\pi n \tilde{R}_7$ for a charge $n$ web, we eventually find
\begin{equation}\label{gcnicenb}
d\tilde{s}_{6}^2 = 2 (\KK_1)_{\eta\etabar} d\eta d\etabar + ds_{\perp}^2 + \delta ds_{\parallel}^2 + \OO(\epsilon^2)~,
\end{equation}
where
\begin{equation}\label{EG}
ds_{\perp}^2= n \tilde{R}_7 \left\{ \frac{1}{2r} \left( dr^2 + r^2 (d\theta^2 + \sin^2{\theta} d\phi^2) \right) + 2r \left[ \frac{d\psi}{n} + \frac{1}{2} \cos{\theta} d\phi \right]^2 \right\}~.
\end{equation}
The remaining $\OO(\epsilon)$ terms in \eqref{gcnicenb} are collected in $\delta ds_{\parallel}^2$.  While their explicit form can be given, the only important feature is that they all have at least one leg along the $\eta$-plane.  Thus, for fixed $\eta$, \eqref{EG} describes the space transverse to the locus at $r = 0$.

When $n=1$, this space is smooth at $r = 0$; after changing variables to $r = 2\rho^2$, we have the metric on flat $\mathbb{R}^4$ in spherical coordinates.  Thus, by including the source locus, $d\tilde{s}_{6}^2$ describes a complete, Ricci-flat K\"ahler manifold.  For $n>1$, \eqref{EG} is the ALE space $A_{n-1} \simeq \mathbb{C}^2/\mathbf{Z}_n$.  The immediate vicinity of the source locus has the form of a $\mathbb{C}$ fibration over $A_{n-1}$.

These statements fit well with one's intuition.  As we zoom in close to a smooth charge $n$ NS5-brane web, it looks approximately like a stack of $n$ coincident straight NS5-branes.  Therefore the T-dual geometry should look approximately like a product of $\mathbb{C}$ and $A_{n-1}$.  The above analysis quantifies these statements.  For example, we can write down the terms in $\delta ds_{\parallel}^2$ and compute higher order corrections to determine quantitatively how the $\mathbb{C}$ fibration over $A_{n-1}$ behaves.  Furthermore, we have learned that the natural radial variable in the transverse space $(y,\lambda,\lambdabar)$ is not the naive one, $y^2 + |\lambda|^2$, but is rather related to this one in an $\eta$-dependent way.

The leading order term in \eqref{gcnicenb}, $2 (\KK_1)_{\eta\etabar} d\eta d\etabar$, is the induced metric on the source locus.  It is not fixed by this perturbative analysis and, in particular, it need not be flat.  In contrast, the near-brane analysis of lower-dimensional brane webs--for example the case of membrane webs \cite{Lunin:2008tf}--reveals that the induced metric on the brane locus is flat.  In this sense NS5-brane webs are the exception to the rule.  The exact solutions to the Monge--Ampere equation we will encounter below, corresponding to particular brane web configurations, do have specific induced metrics on the brane worldvolume which in some cases are non-trivial.  This suggests that the induced metric is fixed when we analytically continue the perturbative solution to an exact one.  It may be fixed by asymptotic boundary conditions, or by matching onto a perturbative solution around a different source locus when there are multiple webs present.

The main technical result of this section is the perturbative solution \eqref{K2gensol} for $\KK$.  In the specific examples below, we will be starting with a Ricci-flat K\"ahler metric, and from it we will construct a function $\KK^{\rm reg}$ via the Legendre transformation and patching procedure described in the previous section, defined on the complement of the fixed lines of a $U(1)$ isometry.  If we can show that $\KK^{\rm reg}$ differs from $\KK$, \eqref{K1K2}, in the vicinity of the fixed lines by a function $k$ with the required properties, \eqref{littlekS}, then we will have constructed $\KK$ on the complement of the fixed lines.  We can then use \eqref{K2gensol} to extend $\KK$ to these fixed lines, thereby identifying the locus of fixed points of the $U(1)$ isometry with the source locus of the Monge--Ampere equation \eqref{sourcedMAS}.

%%%%%%%%%%%%%%%%%%%%%%%%%%%%%%%%%%
\subsection{The case of the singular conifold} \label{singularcase}
%%%%%%%%%%%%%%%%%%%%%%%%%%%%%%%%%%

In \cite{McOrist:2011in} we determined the geometry produced a brane web consisting of two NS5-branes intersecting orthogonally, such that the first brane is located at $y = z^1 = 0$ and the second at $y = z^2 = 0$.  We did this by demanding that the T-dual geometry be given by the singular conifold.  The methods employed there were somewhat roundabout and physical in nature.  Here we present this example in a more streamlined fashion, making use of the machinery developed in the previous two sections.

The first step is to write down the K\"ahler potential corresponding to the Ricci-flat K\"ahler metric on the conifold, as first determined in \cite{Candelas:1989js}.  The conifold is described algebraically by the solution set of the equation
\begin{equation}\label{coneqn}
x u = v w~,
\end{equation}
in $\mathbb{C}^4$.  We identify $(z^1,z^2) = (v,w)$ and introduce two patches: $\UU_+$ where $x \neq 0$ and $(x,v,w)$ are good coordinates; $\UU_-$ where $u \neq 0$ and $(u,v,w)$ are good coordinates.  The K\"ahler potential is $\FF^{\sharp} = \frac{3}{2} L^{2/3} r^{4/3}$, where $r^2 = |x|^2 + |u|^2 + |v|^2 + |w|^2$, and $L$ is an integration constant with dimensions of length controlling the overall ``size'' of the conifold.  Written in terms of independent coordinates in the upper patch,
\begin{equation}\label{Fsing}
\FF^{\sharp}_{+} = \frac{3}{2} L^{2/3} \frac{(|x|^2 + |v|^2)^{2/3}(|x|^2 + |w|^2)^{2/3}}{|x|^{4/3}}~.
\end{equation}
An identical expression holds in the $\UU_-$ patch with $x \to u$.

Let us now introduce the symplectic coordinates $(y,\tilde{x}^7)$ and potential $\KK^{\rm reg}$.  We define the coordinates patchwise via
\begin{align}\label{ysing}
& \UU_+~: \quad c y = \d_{\log{|x|^2}} \FF_{+}^{\sharp} = \frac{L^{2/3} (|x|^4 - |v|^2 |w|^2) }{|x|^{4/3} (|x|^2 + |v|^2)^{1/3} (|x|^2 + |w|^2)^{1/3} } ~,  \cr
& \UU_- ~: \quad cy = -  \d_{\log{|u|^2}} \FF_{-}^{\sharp} = -\frac{L^{2/3} (|u|^4 - |v|^2 |w|^2) }{|u|^{4/3} (|u|^2 + |v|^2)^{1/3} (|u|^2 + |w|^2)^{1/3} } ~,
\end{align}
and
\begin{equation}\label{x7sing}
\UU_+~: \quad \tilde{x}^{7(+)} = c \ {\rm Im}(\log{(x/c)})~, \qquad \UU_- ~: \quad  \tilde{x}^{7(-)} = -c \ {\rm Im}(\log{(u/c)})~.
\end{equation}
On the overlap where both $x,u \neq 0$, we have $x = v w/u$.  Using this relation it is observed that the definitions of $y$ agree so that $y$ is well defined on $\UU_+ \cup \UU_-$, while $\tilde{x}^7$ satisfies
\begin{equation}\label{x7shiftsing}
\tilde{x}^{7(+)} = \tilde{x}^{7(-)} + c \ {\rm Im}( \log{(v w/c^2)})~,
\end{equation}
which is of the form \eqref{x7shift} with $f_1 = c \log{(v w/c^2)}$.  The potential takes the form
\begin{align}\label{Kregsing}
\UU_+~: \quad \KK_{+}^{\rm reg} = \left\{ \FF_{+}^{\sharp} - c y \log{(|x|^2/c^2)} \right\}_{|x| = |x|(y,|v|,|w|)}~, \cr
\UU_-~: \quad  \KK_{-}^{\rm reg} = \left\{ \FF_{-}^{\sharp} + c y \log{(|u|^2/c^2)} \right\}_{|u| = |u|(y,|v|,|w|)}~,
\end{align}
where the subscript denotes that $|x|,|u|$ are to be viewed as functions of $(y,|v|,|w|)$, obtained by inverting \eqref{ysing}.  These definitions differ on the overlap by a transformation of the form \eqref{KT2} with the appropriate $f_1$.

The constant $c$ is not arbitrary.  The coordinates $(y,z^a,\zbar^{\abar})$ are defined such that Ricci-flatness takes the precise form of \eqref{MAnice}.  In terms of the original K\"ahler coordinate system, this means that the determinant of the K\"ahler metric should have the form $g = c^2/8|x|^2$ in the upper patch, for example.  Computing the K\"ahler metric and taking its determinant, one finds that $c$ is given in terms of the length scale $L$ via
\begin{equation}\label{cL}
c = \frac{4}{\sqrt{3}} L~.
\end{equation}

The function $\KK^{\rm reg}$ has been defined on $\UU_+ \cup \UU_-$.  The points not covered are those where $x = u = 0$.  The defining equation for the conifold then implies $v w = 0$, so the remaining points are of the form $x = u = v = 0$, $w$ arbitrary, or $x = u = w = 0$, $v$ arbitrary.  In order to demonstrate that this corresponds to the source locus we must construct the function $\KK$ and show that it has the right singularity structure.

Our discussion in section \ref{Legendre} guarantees that $\KK^{\rm reg}$, \eqref{Kregsing}, satisfies the source free Monge Ampere equation \eqref{MAnice}, and this can also be checked explicitly.  Thus any potential $\KK$ of the form $\KK = \KK^{\rm reg} + k$, where $k$ has the properties \eqref{littlekS} will satisfy the sourced Monge--Ampere equation \eqref{sourcedMAS}.  Therefore in order to demonstrate that the source locus is as claimed, one can proceed as follows. Take a limit that zooms in towards the proposed source locus, using \eqref{Kregsing} to determine the behaviour of $\KK^{\rm reg}$ in the limit. Compare this result with the required near-source form of $\KK$ as given in \eqref{K2gensol}.  If one can define the near-source coordinates $(\eta,\lambda)$ such that the two functions indeed differ by a function $k$ satisfying the required properties, then this will validate the ansatz for the near-source limit, verifying the claimed source structure.

Let us first zoom in towards $w = y = 0$ while keeping $v$ finite.  Specifically we assume
\begin{equation}\label{nbsing}
y = \epsilon \hat{y}~, \qquad w = \epsilon \hat{w}~,
\end{equation}
with $\epsilon$ small and $(\hat{y},v,\hat{w})$ order one.  In this case \eqref{ysing} can be solved perturbatively for $|x|$ or $|u|$ as a function of $y$.  Consistency of \eqref{ysing} with \eqref{nbsing} implies that the leading behaviour of $|x|^2,|u|^2$ is $\OO(\epsilon)$.  We find
\begin{align}\label{xupertsing}
& |x|^2 = \frac{2 L^{1/3} |v|^{2/3}}{\sqrt{3}} \left[ y + \sqrt{ y^2 + \frac{3|v|^{2/3}}{4 L^{2/3}} |w|^2 } \ \right] + \OO(\epsilon^2)~, \cr
& |u|^2 =    \frac{2 L^{1/3} |v|^{2/3}}{\sqrt{3}} \left[ -y + \sqrt{ y^2 + \frac{3 |v|^{2/3}}{4L^{2/3}} |w|^2 } \ \right] + \OO(\epsilon^2)~,
\end{align}
and plugging into $\KK^{\rm reg}$,
\begin{align}\label{Kregnbsing}
& \KK_{\pm}^{\rm reg} = \frac{3}{2} L^{2/3} |v|^{4/3} + c \left\{ D \mp y \log{ \left[ \frac{\sqrt{3} |v|^{2/3}}{8 L^{5/3}} \left( \pm y + D \right) \right]} \right\} + \OO(\epsilon^2)~, \cr
& D = \sqrt{ y^2 + \frac{3|v|^{2/3}}{4 L^{2/3}} |w|^2 }~.
\end{align}

Since the function $k$ can not affect the order one term corresponding to the induced metric on the locus, we must, according to \eqref{K1K2}, identify
\begin{equation}\label{inducedsing}
(\KK_1)_{\eta\etabar} d\eta d\etabar = \frac{2 L^{2/3}}{3 |v|^{2/3}} dv d\vbar~.
\end{equation}
Once we choose a coordinate $\eta$, the coordinate $\lambda$ is determined by the requirement that the Jacobian for the change of variables $(v,w) \to (\eta,\lambda)$ have unit determinant.  One choice is simply
\begin{equation}\label{nbcoord1}
(\eta,\lambda) = \left( v + \OO(\epsilon),w + \OO(\epsilon^2) \right)  \quad \Rightarrow \quad (\KK_1)_{\eta\etabar} = \frac{2 L^{2/3}}{3 |\eta|^{2/3}}~.
\end{equation}
However, we observe that in this example $\KK_1$ is the mod-squared of a holomorphic function, so we may choose coordinates such that the induced metric on the source locus is (locally) flat:
\begin{equation}\label{nbcoord2}
\eta = \sqrt{3} L^{1/3} v^{2/3} + \OO(\epsilon) ~, \qquad \lambda = \frac{\sqrt{3}}{2 L^{1/3}} v^{1/3} w + \OO(\epsilon^2)~,
\end{equation}
implying $(\KK_1)_{\eta\etabar} = \half$.  With either choice of coordinates $(\eta,\lambda)$ we find that \eqref{Kregnbsing} can be put in the form
\begin{align}\label{Kregnbsing2}
& \KK_{\pm}^{\rm reg} = \KK_1 + c \left\{ D \mp y \log{\left[ \sqrt{2 (\KK_1)_{\eta\etabar}} \left( \pm y + D \right) \right]} \right\} + y h_1(\eta,\etabar)  + \OO(\epsilon^2)~, \cr
& D = \sqrt{ y^2 + \frac{1}{2 (\KK_1)_{\eta\etabar}}  |\lambda|^2 }~,
\end{align}
for a particular $h_1(\eta,\etabar)$.  We can use this expression to compute the leading behaviour of the warp factor, $e^{-3A} = 4 \det{(\d \delbar \KK^{\rm reg})}$.  Upon comparing the result with \eqref{warpnb}, we determine a relation between $c$ and the charge: $c = \tilde{Q}_0/2\pi$.  If the conifold is to be T-dual to a charge one brane web, then $\tilde{Q}_0 = 2\pi \tilde{R}_7$, implying $c = \tilde{R}_7$.  Note that this also gives the coordinate $\tilde{x}^7$, as defined in \eqref{xitoy}, the right periodicity.  Furthermore this implies a relationship between the size parameter of the conifold and the asymptotic radius of the $\tilde{x}^7$ circle:
\begin{equation}\label{throatsize}
c = \tilde{R}_7 \quad \Rightarrow \quad L = \frac{\sqrt{3}}{4} \tilde{R}_7~.
\end{equation}

With $\KK^{\rm reg}$ in the form \eqref{Kregnbsing2} and $\KK$ given by \eqref{K1K2}, \eqref{K2gensol}, it is easy to check that the difference $k = \KK^{\rm reg} - \KK$ does satisfy the conditions \eqref{littlekS}.  The partial derivatives of $\KK^{\rm reg}$ have exactly the same form as those of $\KK$, \eqref{Kpartials}, but without the singular terms in the $y$-derivatives; hence \eqref{littlekS} follows.  This demonstrates that the coordinate system $(\eta,\lambda)$, as defined in \eqref{nbcoord1} or \eqref{nbcoord2}, properly describes the source locus.  It is located at $\lambda = y = 0$ and parameterised by $\eta$.  For either choice \eqref{nbcoord1} or \eqref{nbcoord2}, this corresponds to $y = w = 0$, $v$ finite.  The different coordinate choices merely correspond to the freedom to make holomorphic coordinate reparameterisations which have unit determinant and preserve the boundary condition that the source locus is at $\lambda = 0$.

Since the exact solution \eqref{Kregsing} for $\KK^{\rm reg}$ is symmetric under the exchange $v \leftrightarrow w$, it is clear that we can carry out an identical analysis to demonstrate that $y = v = 0$, $w$ finite, corresponds to a source locus as well.  The only part of the geometry our analysis has not covered is the infinitesimal neighbourhood around $y = v = w = 0$.  This point corresponds to the conifold singularity on the geometric side and the intersection point of the two NS5-branes on the brane web side.  The classical geometry is singular at this point and we will not consider it further here.

In conclusion, we have found the exact potential, $\KK^{\rm reg}$, that describes both the conifold geometry through \eqref{gcnice2} and the brane web geometry via \eqref{NSfive} (smeared on the $x^7$ circle).  $\KK^{\rm reg}$ is defined everywhere except at the locus $y = v = 0$ and $y = w = 0$.  By comparing the form of $\KK^{\rm reg}$ with the general form of a solution $\KK$ to the sourced Monge--Ampere equation \eqref{sourcedMAS}, we identified this locus with the position of the brane web. The conifold geometry extends smoothly to the locus (except at the intersection point), whereas the brane web geometry is singular at the locus.  This is natural since the brane web is composed of fundamental sources for the metric, dilaton, and Neveu--Schwarz three-form flux.  With a convenient choice of coordinates one can write down the brane web geometry in completely explicit form.  We refer the reader to \cite{McOrist:2011in} for details.

%%%%%%%%%%%%%%%%%%%%%%%%%%%%
%%%%%%%%%%%%%%%%%%%%%%%%%%%%
\section{T-dualising the resolved conifold}
%%%%%%%%%%%%%%%%%%%%%%%%%%%%
%%%%%%%%%%%%%%%%%%%%%%%%%%%%

The resolved conifold is a deformation of the singular conifold that preserves Ricci-flatness of the metric. Intuitively, one can think of it as resolving the singularity by an $S^2$. The Ricci-flat metric for the resulting space is known and admits a $U(1)$ isometry. T-dualising in this direction results in a pair of separated NS5-branes, which we construct explicitly in this section. We show how, for example, the parameter describing the resolution (size of the $S^2$) maps to the separation of the NS5-branes.

%%%%%%%%%%%%%%%%%%%%%%%%%%%%%%%
\subsection{A rapid review of the resolved conifold}
%%%%%%%%%%%%%%%%%%%%%%%%%%%%%%%

Let $(Z_1,Z_2)$ be the homogenous coordinates for a $\P^1$ so that they are not both zero. Then the resolved conifold may be described as the solution space to the equation
\begin{equation}
 \WW \begin{pmatrix} Z_1 \\ Z_2 \end{pmatrix}  = \begin{pmatrix}0 \\ 0\end{pmatrix}~,\quad {\rm with} \quad \WW = \begin{pmatrix}v & x \\ u & w \end{pmatrix}~.\label{eq:resdef}
\end{equation}
As $(Z_1,Z_2) \neq (0,0)$, the conifold equation, $x u - v w = 0$, is satisfied and ${\rm rank}~\mathcal{W} \le 1$. If ${\rm rank}~\mathcal{W} = 1$ then we can solve for $Z_1$ in terms of $Z_2$, and this specifies a unique point on $\P^1$. If ${\rm rank}~\mathcal{W} = 0$ then $(Z_1,Z_2)$ parameterise a full $\P^1$. Intuitively, one can think of the solution space of (\ref{eq:resdef}) to match that of the conifold except at the conifold singularity where the singularity is  replaced by a $\P^1$.

In fact, it follows from (\ref{eq:resdef}) that the resolved conifold, $\BBh$, is the total space of an $\OO(-1) \oplus \OO(-1)$ fibration over $\P^1$:
\begin{equation}
 \begin{array}{ccc}
  \OO(-1) \oplus \OO(-1) & \rightarrow & \BBh \\
  && \downarrow \\
  && \P^1
 \end{array}\label{eq:bundle}
\end{equation}
It is instructive to see how this comes about in detail. To that end, we begin by introducing $H_S$, $H_N$, the usual patches covering $\P^1$ defined by stereographic projection from the south and north poles.  In each patch we define a complex coordinate,
\begin{equation}
 H_N: \quad Z = Z_2/Z_1~, \qquad H_S : \quad Y =  Z_1/Z_2~,
\end{equation}
with $Z = 1/Y$ on the overlap $H_S \cap H_N$. Recall, the coordinate transformation $Z = \tan(\frac{\theta}{2})\, e^{-i\phi}$ maps the Fubini--Study metric to the round $S^2$ metric:
\begin{equation}
 \frac{4 dZ d\Zbar}{(1+|Z|^2)} = d\theta^2 + \sin^2 \theta d \phi^2~.\label{eq:FubiniStudy}
\end{equation}
For $(Z_1,Z_2)\in H_N$ we can parameterise solutions of (\ref{eq:resdef}) by letting $v = - Z x$ and $u = - Z w$ so that
\begin{equation}
 \WW = \begin{pmatrix}
        -Zx & x \\
        -Z w & w
       \end{pmatrix}~.
\label{eq:WforHS}
\end{equation}
Now $(x,w)$ are coordinates for the $\OO(-1)\oplus \OO(-1)$ fibre over $H_N$. We denote this trivialisation by $\HH_N = \OO(-1)\oplus \OO(-1)\times H_N$ which is a patch in the total space $\BBh$.

Which solutions of (\ref{eq:resdef}) are not accounted for by (\ref{eq:WforHS})?  Precisely the ones of the form
\begin{equation}
 \WW = \begin{pmatrix}
        * & 0\\
	* & 0
       \end{pmatrix}~,
\label{eq:WHSmisses}
\end{equation}
where ``$*$'' represents nonzero entries. These solutions occur when $(Z_1,Z_2)\in H_S$ with
\begin{equation}
 \WW = \begin{pmatrix}
        v & - Y v \\
        u & - Y u
       \end{pmatrix}~,
\label{eq:WforHN}
\end{equation}
where now $(u,v)$ parameterise $\OO(-1)\oplus \OO(-1)$ over $H_S$. As for above we denote $\HH_S =  \OO(-1)\oplus \OO(-1)\times H_S$. On the overlap $\HH_N \cap \HH_S$  the respective coordinates of (\ref{eq:WforHS}) and (\ref{eq:WforHN}) are related by
\begin{equation}
 (v,u; Y) = (-Zx, -Z w; 1/Z)~,\label{eq:transitions}
\end{equation}
and this is the transition function defining an $\OO(-1) \oplus \OO(-1)$ bundle over $\P^1$.  The patches $\HH_S$ and $\HH_N$ cover $\BBh$.  The solution space of (\ref{eq:resdef}) is indeed described by the bundle (\ref{eq:bundle}) and we are done.

The Ricci-flat K\"ahler metric is constructed in \cite{Candelas:1989js} as follows. Denote the K\"ahler potential $\FFh$ and suppose we are in the patch $\HH_N$ with coordinates $(w,x;Z)$. Then, requiring the metric to be homogenous under patching leads one to the ansatz
\begin{equation}
 \FFh = \FF(r^2;a) + 4 a^2 \log{(1 + |Z|^2)}~,
 \label{eq:Fresansatz}
\end{equation}
where $r^2 = \tr( \WW^{\dag} \WW)$ and $\FF(r^2;a)$ is a function to be determined. Label the coordinates $z^\alpha = (w,x;Z)$; then the metric takes the form
\begin{equation}
d\hat{s}^2 = 2 \left[ ( \d_\alpha \delbar_{\bar\beta} r^2) \FF' + (\d_\alpha r^2) (\delbar_{\bar\beta} r^2) \FF'' \right] d\zeta^\alpha d{\bar\zeta}^{\bar\beta} + 4 a^2  \frac{dZ d\Zbar}{(1 + |Z|^2)^2} ~,
 \label{eq:resKahmet}
\end{equation}
where prime denotes differentiation with respect to $r^2$. There are two interesting limits: $a\to0$ and $r\to0$. Under the former, we recover the singular conifold and this allows us to identify $\FFh(r^2;0) = \FF(r^2;0) = \FF^{\sharp}(r^2)$, the K\"ahler potential of the singular conifold. The latter limit corresponds to zooming in towards the resolved singularity. Indeed, the first two terms in (\ref{eq:resKahmet}) vanish as $r \to 0$, and therefore, in accordance with (\ref{eq:FubiniStudy}), we get the round metric on an $S^2$ with radius $a$; the parameter $a$ is referred to as the resolution parameter.

After turning the crank slowly but surely, the Ricci-flatness condition $R_{\alpha{\bar\beta}} = \d_\alpha \delbar_{\bar\beta} g = 0$, with $g = \det g_{\alpha\bar\beta}$, reduces to the simple equation
\begin{equation}
 \gamma' \gamma \left( \gamma + 2 a^2 \right) = \frac{2}{3} L^2 r^2~, \qquad {\rm where} \quad \gamma(r^2;a) \equiv r^2 \FF'(r^2;a)~,
\label{eq:resRicciflat}
\end{equation}
and $L$ is an integration constant.  The limit $a\to0$ also implies $L$ is to be identified with the integration constant introduced in \cite{McOrist:2011in}, related to the asymptotic radius of the $\tilde{x}^7$ circle, ${\tilde R}_7 = \frac{4 L}{\sqrt{3}}$, in the asymptotically flat geometry T-dual to the intersecting NS5-branes.

Equation (\ref{eq:resRicciflat}) can be integrated, giving
\begin{equation}
 \gamma^3 + 3 a^2 \gamma^2 - L^2 r^4 = 0~,
\label{eq:gammacubic}
\end{equation}
where another integration constant has been set to zero such that (\ref{eq:resKahmet}) reduces to the round metric on $S^2$ when $r \to 0$.  It is easy to show that for all $r > 0$ this cubic has a unique positive real root $\gamma_0$, and as $r \to 0$, $\gamma_0  \to 0$. One can find an explicit expression for $\FF(r^2;a)$ by writing out this root,
\begin{align}
 \gamma_0 =&~ a^2 ( -1 + \alpha^{-1/3} + \alpha^{1/3} )~, \qquad {\rm where} \cr
\alpha =&~  \xi^2 - 1 + \sqrt{ \xi^4 - 2 \xi^2}~, \qquad \xi \equiv \frac{L r^2}{2 a^3}~,
\label{eq:cubicroot}
\end{align}
 and then integrating  with respect to $r^2$, as dictated by (\ref{eq:resRicciflat}).  After dropping an irrelevant integration constant, we find
\begin{equation}
\FF(r^2;a) = \frac{3}{2} \left\{ \gamma_0 - a^2 \log{\left[ 3 + \frac{\gamma_0}{a^2} \right]} \right\}~.
 \label{eq:Fresult}
\end{equation}

As shown in \cite{PandoZayas:2000sq}, the metric has an expression in a conical form given by introducing the angular coordinates,
 \begin{align}
& x =  r \cos{ \frac{\theta_1}{2} } \cos{ \frac{\theta_2}{2} } e^{ \frac{i}{2} (\psi + \phi_1 + \phi_2) }~, \cr
& w = r \sin{ \frac{\theta_1}{2} } \cos{ \frac{\theta_2}{2} } e^{ \frac{i}{2}(\psi - \phi_1 + \phi_2) }~, \cr
& Z = \tan{ \frac{\theta_2}{2} } e^{-i \phi_2}~.
 \label{eq:TPZcov}
 \end{align}
 Changing the radial variable from $r$ to $\rho^2=3\gamma$ and using
 \begin{equation}
 \gamma' = \frac{2 L^2 r^2}{3 \gamma (\gamma + 2 a^2)}~,
  \label{eq:gammaprime}
 \end{equation}
 one can show that the metric takes the form
 \begin{align}
& d\hat{s}^2 = \kappa^{-1}(\rho) d\rho^2 + \frac{1}{6} \rho^2 \left( d\theta_{1}^2 + \sin^2{\theta_1} d\phi_{1}^2 \right) + \frac{1}{6} (\rho^2 + 6 a^2 ) \left( d\theta_{2}^2 + \sin^2{\theta_2} d\phi_{2}^2  \right)  + \cr
& \qquad ~ + \frac{1}{9} \kappa(\rho) \rho^2 \left[ d\psi + \cos{\theta_1} d \phi_1 + \cos{\theta_2} d\phi_2 \right]^2 ~,
 \label{eq:TPZmet}
 \end{align}
 where
 \begin{equation}
  \kappa(\rho) \equiv \frac{ \rho^2 + 9 a^2 }{\rho^2 + 6 a^2}~.
  \label{eq:kappadef}
 \end{equation}
 In the $a \rightarrow 0$ limit this metric reduces to the metric on the singular conifold, presented as a cone over the Einstein space $T^{1,1}$.  On the other hand, when $\rho \rightarrow 0$, there is a two-sphere of radius $a$, corresponding to the $\P^1$ of the resolution.

This whole discussion has been carried out in the patch $\HH_N \subset \BBh$, where $Z_1 \neq 0$.  This patch covers everything except for the $\C \oplus \C$ fibre over the south pole of $\P^1$, where $Z$ is no longer a good coordinate.  To describe the neighbourhood of the south pole (and the fibres over it), we should use the coordinate system $(u,v; Y)$, where $Y = Z_1/Z_2$.  Doing so, one derives an identical expression to (\ref{eq:resKahmet}), but with $Z \to Y$ and where $r^2$ should be set equal to $r^2 = (|v|^2 + |u|^2)(1 + |Y|^2)$.  Upon making the change of variables
\begin{align}
& u =- r \sin{ \frac{\theta_1}{2} } \sin{ \frac{\theta_2}{2} } e^{ \frac{1}{2} i (\psi - \phi_1 - \phi_2) }~, \cr
&  v = - r \cos{ \frac{\theta_1}{2} } \sin{ \frac{\theta_2}{2} } e^{ \frac{1}{2} i (\psi + \phi_1 - \phi_2) }~, \cr
& Y = \cot{ \frac{\theta_2}{2} } e^{i \phi_2}~,
 \label{eq:TPZcov2}
 \end{align}
one again recovers \eqref{eq:TPZmet}.  Note that \eqref{eq:TPZcov}, \eqref{eq:TPZcov2} are obtained as in \cite{Candelas:1989js} by writing $\WW = L \WW_0 R^{\dag}$, with $\WW_0 = \half (\sigma^1 + i \sigma^2)$, and $L,R \in SU(2)$ expressed in terms of Euler angles.

%%%%%%%%%%%%%%%%%%%%%%%%%%%%%%%%%%%%%%%%%%%
\subsection{Legendre transformations around the source locus}
%%%%%%%%%%%%%%%%%%%%%%%%%%%%%%%%%%%%%%%%%%%

We now write the resolved conifold metric in the form (\ref{gcnice2}), by making use of the Legendre transformation reviewed in section \ref{Legendre} to exchange the K\"ahler coordinate $x$ or $u$ with symplectic coordinates $(y,\tilde{x}^7)$. We perform this transformation away from the source locus, where $\KK^{\rm reg}$ satisfies the source-free Monge--Ampere equation. To do this, we first need to refine our atlas of patches covering the resolved conifold. This is so because the patches $\HH_S, \HH_N$ cover $\BBh$, and in particular, include the source locus. We then extend $\KK$ to the source locus in such a way that it satisfies the sourced equation.

As in the case of the singular conifold we identify $(z^1, z^2) = (v,w)$ and find that it is useful to consider the patches
\begin{equation}
 \UU_+ : \quad x \neq 0~, \qquad \UU_- : \quad u \neq 0~,
 \label{eq:pmpatch}
\end{equation}
where it is straightforward to observe that $\UU_+ \subset \HH_N$, and $\UU_- \subset \HH_S$. We can write the K\"ahler potential (\ref{eq:Fresansatz}) in the patch $\UU_+$ by substituting $Z=-v/x$ and using complex coordinates $(x,v,w)$:
\begin{align}
\UU_+ : \qquad & \FFh_+(|x|,|v|,|w|) = \FF(r^2;a) + 2 a^2 \log{\left(1 + \frac{|v|^2}{|x|^2} \right)}~, \qquad {\rm with} \cr
& r^2 = \frac{(|x|^2 + |v|^2) (|x|^2 + |w|^2)}{|x|^2}~.\label{eq:FresUp}
\end{align}
Similarly, in $\UU_-$ we use $Y=-w/u$ to write
\begin{align}
\UU_- : \qquad & \FFh_-(|u|,|v|,|w|) = \FF(r^2;a) + 2 a^2 \log{\left(1 + \frac{|w|^2}{|u|^2} \right)}~, \qquad {\rm with} \cr
& r^2 = \frac{(|u|^2 + |v|^2) (|u|^2 + |w|^2)}{|u|^2}~.
\label{eq:FresUm}
\end{align}
Now introduce the symplectic coordinates $(y,\tilde x^7)$ via the Legendre transformation analogous to (\ref{ysing}):
\begin{align}
 \UU_+~: \qquad c(y - y_+) &= |x|^2 \d_{|x|^2} \FFh_+ \cr
 &= \frac{ |x|^4 - |v|^2 |w|^2 }{(|x|^2 + |v|^2)(|x|^2 + |w|^2)} \gamma(r^2;a) - \frac{2 a^2 |v|^2}{|x|^2 + |v|^2}~,\label{eq:yUp}  \\
 \UU_-~: \qquad c(y - y_-) &= - |u|^2 \d_{|u|^2} \FFh_- \cr
& =  - \frac{ |u|^4 - |v|^2 |w|^2 }{(|u|^2 + |v|^2)(|u|^2 + |w|^2)} \gamma(r^2;a) + \frac{2 a^2 |w|^2}{|u|^2 + |w|^2}~.\label{eq:yUm}
\end{align}
where $\gamma$ is given by the solution of the cubic (\ref{eq:gammacubic}) and we have used (\ref{eq:FresUp}) and (\ref{eq:resRicciflat}).  The relative sign here is consistent with the relationship between $x,u$ on $\UU_+ \cap \UU_-$ and the general guidelines discussed around \eqref{yKTs}.  The isometry direction $\tilde x^7$ is identified as in (\ref{x7sing}):
\begin{equation}
\label{eq:yxsevenp}
\UU_+ :  \quad {\tilde x}^{7(+)} = c\, {\rm Im}(\log(x/c))~,\qquad \UU_-:  \quad{\tilde x}^{7(-)} = -c\, {\rm Im}(\log(u/c))~.
\end{equation}
The relationship $c = 4L/\sqrt{3}$ is again obtained by requiring the determinant of the K\"ahler metric to have the correct normalisation (see discussion following \eqref{detgKahler}).

What points in $\BBh$ are missed by $\UU_+ \cup \UU_-$? They are clearly characterised by $x=u=0$ and may be further subdivided into three types:
\begin{enumerate}
 \item $x=u=v=0$ with $w\neq0$. This implies $Z=0$ fixing us to the north pole of the $\P^1$;
\item $x=u=w=0$ with $v\neq0$. This implies $Y=0$ fixing us to the south pole of the $\P^1$;
\item $x=u=v=w=0$ with $[Z_1,Z_2]$ unconstrained and parameterising the $\P^1$ of the small resolution.
\end{enumerate}
Our intuition from the singular conifold is that the first two types of points correspond to the source locus.  This will be confirmed in the next section; here for convenience we introduce some standard nomenclature \cite{Giveon:1998sr}, referring to points of type 1 as the $NS'$ locus and points of type 2 as the $NS$ locus.  Points of type 3 are not on the source locus, and we need at least one more coordinate patch to cover the complement of the source locus in $\BBh$.  Before considering this however, it will help to consider some limits of (\ref{eq:yUp})-(\ref{eq:yUm}).

Let us focus on (\ref{eq:yUp}) by working in the $\UU_+$ patch.  Firstly, suppose we approach a point on the $NS'$ locus by taking the limit  $|x|\rightarrow 0$ with $v=0$ in (\ref{eq:yUp}):
\begin{align}
 c(y - y_+) \displaystyle\bigg|_{v = 0} &=~ \frac{|x|^2}{|x|^2 + |w|^2} \gamma(r^2;a) \cr
& \longrightarrow~  0~, \qquad\qquad {\rm as} \quad |x| \to 0~.
\label{eq:yUpvzero}
\end{align}
Secondly, let's zoom in on the $NS$ locus by letting $w = 0$ and assuming $v \neq 0$:
\begin{align}
c(y - y_+) \displaystyle\bigg|_{w = 0} &=~ \frac{|x|^2}{|x|^2 + |v|^2} \gamma(r^2;a) - \frac{2 a^2 |v|^2}{|x|^2 + |v|^2} \cr
& \longrightarrow ~ -2 a^2~, \qquad\qquad {\rm as} \quad |x| \to 0~.
 \label{eq:yUpwzero}
\end{align}
If our intuition about the location of the branes is correct, then this shows that the $NS'$ and $NS$ are separated in $y$ by an amount
\begin{equation}
 \Delta y = \frac{2 a^2}{c}~.
 \label{eq:Deltay}
\end{equation}
At this point it is convenient to choose $y_{\pm} = \pm a^2/c$ as the $y$-location of the $NS'$ and $NS$ respectively.

The parameter $y$ takes values in different ranges depending on the value of $(v,w)\in\UU_+$:
\begin{itemize}
 \item if $v\neq0$ and $w \neq 0$, then $y\in(-\infty, \infty)$;
 \item  if $v=0$ and $w$ arbitrary, then $y\in(a^2/c,\infty)$;
\item if $w=0$ and $v\ne0$, then $y\in(-a^2/c, \infty)$.
\end{itemize}
Inuitively we expect $y\in(-\infty,\infty)$ for all values of $v,w$ and the missing values of $y$ must be accounted for using other patches. The definition of $y$ in the $\UU_-$ patch is given by analysing (\ref{eq:yUm}), and there we find the following:
\begin{itemize}
 \item if $v\neq0$ and $w \neq 0$, then $y \in (-\infty,\infty)$;
\item if  $v=0$ and $w\ne0$, then $y \in (-\infty, a^2/c)$;
\item  if $ w = 0$ and $v$ arbitrary, then $y \in (-\infty, -a^2/c)$.
\end{itemize}

It must be that the locus in which $v = w = 0$ and $y\in(- a^2/c, a^2/c)$ is the $\P^1$ at $r = 0$, and furthermore this locus is not contained in  $\UU_+ \cup \UU_-$. Consequently,  we define two additional patches:
\begin{equation}
 \UU_Z : \quad Z \neq 0~, \qquad \UU_Y : \quad Y \neq 0~.
\label{eq:UZUY}
\end{equation}
As sets, $\UU_Y = \UU_Z = \HH_S \cap \HH_N$, but we will distinguish $\UU_Z$, $\UU_Y$ by the coordinate charts we use on each of them.  The chart we use on $\UU_Z \subset \HH_N$ is given by eliminating $x$ via $x = -v/Z$, such that $(v,w,Z)$ are our coordinates.   On the other hand, on $\UU_Y \subset \HH_S$ we eliminate $u$ in favor of $u = -w/Y$, such that $(v,w,Y)$ are our coordinates.  The reason for making this distinction is that it will aid in the global analysis: the limits $Z \to 0$ and $Y \to 0$ are different.

The K\"ahler potential on each of these patches takes the following form:
\begin{align}
\UU_Z : \qquad & \FFh_{Z}(|Z|,|v|,|w|) = \FF(r^2; a) + 2 a^2 \log{ \left( 1 + |Z|^2 \right)},~ \qquad {\rm with} \cr
& r^2 = (1 + |Z|^2) \left( |w|^2 + \frac{ |v|^2}{|Z|^2} \right)~,
 \label{eq:FresZ}
\end{align}
and
\begin{align}
\UU_Y : \qquad & \FFh_{Y}(|Y|,|v|,|w|) = \FF(r^2; a) + 2 a^2 \log{ \left( 1 + |Y|^2 \right)},~ \qquad {\rm with} \cr
& r^2 = (1 + |Y|^2) \left( |v|^2 + \frac{ |w|^2}{|Y|^2} \right)~.
 \label{eq:FresY}
\end{align}
We may apply the Legendre transformation procedure, with $\log{|Z|^2}$, or $\log{|Y|^2}$ respectively, playing the role of the coordinate dual to $y$.  We have
\begin{align}
\UU_{Z} &: \quad c y - a^2 = - |Z|^2 \d_{|Z|^2} \FFh_{Z}~, \qquad {\tilde x}^{7(Z)} = -c \phi_Z~, \qquad {\rm and} \cr
\UU_Y &: \quad c y + a^2 =  |Y|^2 \d_{|Y|^2} \FFh_{Y}~, \qquad {\tilde x}^{7(Y)} = c \phi_Y~.
\label{eq:yxsevenZY}
\end{align}
The signs on the right-hand sides of these expressions were determined by the relationship between $x,Z$ on $\UU_+ \cap \UU_Z$ and $u,Y$ on $\UU_- \cap \UU_Y$.  Writing out these definitions of $y$ gives
\begin{align}
 & \UU_Z : \quad cy - a^2 = - \frac{ |Z|^4 |w|^2 - |v|^2 }{|Z|^4 |w|^2 + |v|^2 + |Z|^2 (|v|^2 + |w|^2)} \gamma(r^2 ;a) - \frac{2 a^2 |Z|^2}{1 + |Z|^2}~, \cr
& \UU_Y: \quad cy + a^2 =  \frac{ |Y|^4 |v|^2 - |w|^2 }{|Y|^4 |v|^2 + |w|^2 + |Y|^2 (|v|^2 + |w|^2)} \gamma(r^2 ;a) + \frac{2 a^2 |Y|^2}{1 + |Y|^2}~.
\label{eq:yUZY}
\end{align}
Using the relation $|Z| = 1/|Y|$, one finds that these two definitions of $y$ agree. Furthermore it easy to check using $|Z| = |v|/|x|$ that the definition on $\UU_Z$ agrees with the one on $\UU_+$, and using $|Y| = |w|/|u|$ that the definition on $\UU_Y$ agrees with the one on $\UU_-$.  It follows that definitions agree on all overlaps.  As long as one of $v,w$ is not zero, the range of $y$ determined from \eqref{eq:yUZY} agrees with the above.  When $v = w = 0$ it is easy to check that $y \in (-a^2/c, a^2/c)$.

Putting all of the pieces together, we have the following picture.
\begin{itemize}
\item The image of the $NS'$-brane is parameterised by $w$ in the fibre above the north pole of $\P^1$ and is located at $y=a^2/c$.  For $y > a^2/c$ and $v = w = 0$, the $y$-${\tilde x}^7$ cylinder is identified with $\C_{x}^\ast$.
\item The image of the $NS$-brane is parameterised by $v$ in the fibre above the south pole of $\P^1$ located at $y = -a^2/c$.  For values of $y < -a^2/c$, the $y$-${\tilde x}^7$ cylinder at $v = w= 0$ is identified with $\C_{u}^\ast$.
\item For $y \in (-a^2/c, a^2/c)$, the $y$-${\tilde x}^7$ cylinder corresponds to $\C_{Z}^\ast \simeq \C_{Y}^\ast$.  The north pole corresponds to $Z = 0$ and the south pole to $Y = 0$.
\item    The four patches $\UU_\pm$, $\UU_Z$, $\UU_Y$ together cover everything except for the source locus, and we have given the coordinate $y$ a global definition across all of the patches.
\end{itemize}

The Legendre transformation also tells us how to construct the brane potential, $\KK^{{\rm reg}}$, in each patch:
\begin{align}
& \UU_+ ~: \quad \KK^{{\rm reg}}_+(y) = \left[ \FFh_+(|x|) - (cy - a^2) \log{(|x|^2/c^2)} \right]_{|x|^2 = |x|^2(y)} ~, \cr
& \UU_Z ~: \quad \KK_{Z}^{\rm reg}(y) = \left[ \FFh_Z(|Z|) + (cy - a^2) \log{|Z|^2} \right]_{|Z|^2 = |Z|^2(y)} ~, \cr
& \UU_Y ~: \quad \KK_{Y}^{\rm reg}(y) = \left[ \FFh_Y(|Y|) - (cy + a^2) \log{|Y|^2} \right]_{|Y|^2 = |Y|^2(y)} ~, \cr
& \UU_- ~: \quad \KK_{-}^{\rm reg}(y) = \left[ \FFh_-(|u|) + (cy + a^2) \log{(|u|^2/c^2)} \right]_{|u|^2 = |u|^2(y)} ~.
 \label{eq:resKdef}
\end{align}
We have suppressed $(|v|,|w|)$ in the arguments of all functions, and the subscripts are to indicate that $|x|^2, |u|^2, |Y|^2, |Z|^2$ are to be viewed as functions of $(y,|v|,|w|)$ obtained by inverting (\ref{eq:yUp}), (\ref{eq:yUm}) or (\ref{eq:yUZY}) as appropriate.   $\KK^{\rm reg}$ solves the source-free Monge--Ampere equation everywhere away from the locus.  In the next section we study its behaviour as we approach the source locus, and show that it is consistent with a $\KK$ that solves the sourced Monge--Ampere equation.  This will complete the demonstration that the resolved conifold is T-dual to an NS5-brane system whose brane locus is identified with the source locus above.

%%%%%%%%%%%%%%%%%%%%%%%%%%%%%%%%%%
\subsection{Extending $\KK$ to the source locus}
%%%%%%%%%%%%%%%%%%%%%%%%%%%%%%%%%%

We study the $NS$ locus in detail; the analysis for the $NS'$ locus is nearly identical.  The $NS$ locus is described by $y+a^2/c = w=0$, and corresponds to $\HH_S \setminus (\UU_- \cup \UU_Y)$. We may approach the locus in two ways. The first is from within the set $\UU_-$ by taking the limit $y + a^2/c \to 0^-$ (that is, send $u,w \to 0$). The second is from within the set $\UU_Y$ by taking a limit $y + a^2/c \to 0^+$ (that is, send $Y,w \to 0$). In each case we solve for $|u|$ or $|Y|$ as a function of $(y,|v|,|w|)$ via (\ref{eq:yUm}) or (\ref{eq:yUZY}) respectively.

Let's start with $\UU_-$ by setting
\begin{equation}
y + a^2/c = \epsilon {\hat y}~, \qquad w = \epsilon {\hat w}~,
 \label{eq:smallyv}
\end{equation}
where $\epsilon\ll1$ and ${\hat y}, {\hat w}, v$ are $\OO(1)$. From (\ref{eq:yUm}) we see the leading order scaling behaviour of  $|u|^2$ is $\OO(\epsilon)$.  Writing $|u|^2 = \epsilon (|u_0|^2 + \cdots)$ and solving \eqref{eq:yUm} perturbatively, we find
\begin{equation}
|u_0|^2 = \frac{c |v|^2}{2 \gamma_{0v}} \left[ - \hat{y} + \sqrt{ {\hat y}^2 +\frac{ |{\hat w}|^2}{2 \gamma_{0v}'}  } \ \right]~,
 \label{eq:uroot}
\end{equation}
with $\gamma_{0v} = \gamma(|v|^2;a)$ being the solution to $\gamma_{0v}^2 (\gamma_{0v} + 3 a^2) = L^2 |v|^4$ and $\gamma_{0v}' = \del_{|v|^2} \gamma_{0v}$.  In order to obtain this result we made use of both \eqref{eq:gammaprime} and the relation $c = 4L/\sqrt{3}$.  We can plug \eqref{eq:uroot} into $\KK_{-}^{\rm reg}$, \eqref{eq:resKdef}, and expand in $\epsilon$.  After some manipulation we find that the result can be expressed as
\begin{align}\label{Kresm}
\KK_{-}^{\rm reg} =&~ \FF(|v|^2;a) + c \left\{ D+ (y+a^2/c) \log{\left[ \frac{\sqrt{2 \gamma_{0v}'}}{c} \left( -(y+a^2/c) + D  \right)\right]} \right\}  + \cr
&~ + \half (cy + a^2) \log{\left[ \frac{|v|^2}{c^2} \left(1 + \frac{2 a^2}{\gamma_{0v}} \right) \right]}  + \OO(\epsilon^2)~, \qquad \textrm{where} \cr
D =&~ \sqrt{ (y+a^2/c)^2 +\frac{ |w|^2}{2 \gamma_{0v}'}  }~.
\end{align}
It is straightforward to repeat this calculation in the $\UU_Y$ patch.  We solve \eqref{eq:yUZY} perturbatively in $\epsilon$ for the leading behaviour of $|Y|^2$ and then plug the result into $\KK_{Y}^{\rm reg}$.  We find a result that differs from \eqref{Kresm} by a couple appropriately placed signs:
\begin{align}\label{KresY}
\KK_{Y}^{\rm reg} =&~ \FF(|v|^2;a) + c \left\{ D - (y+a^2/c) \log{\left[ \frac{\sqrt{2 \gamma_{0v}'}}{c} \left( (y+a^2/c) + D  \right)\right]} \right\}  + \cr
&~ + \half (cy + a^2) \log{\left[ \frac{|v|^2}{c^2} \left(1 + \frac{2 a^2}{\gamma_{0v}} \right) \right]}  + \OO(\epsilon^2)~.
\end{align}

If we take our near-source coordinates to be
\begin{equation}\label{nbcoordres}
(\eta,\lambda) = \left( v + \OO(\epsilon), w + \OO(\epsilon)^2 \right)~,
\end{equation}
and identify $\KK_1(\eta,\etabar) = \FF(|\eta|^2;a)$, then \eqref{Kresm} and \eqref{KresY} have the general form
\begin{align} \label{KregNSbrane}
\KK^{\rm reg} =&~ \KK_1 + c \left\{ D \mp (y -y_-) \log{\left[ \sqrt{2 (\KK_1)_{\eta\etabar}} \left( \pm (y -y_-) + D\right) \right]} \right\} + \cr
& +  (y -y_-) h_1(\eta,\etabar)~, \qquad \textrm{with} \cr
D =&~ \sqrt{ (y-y_-)^2 + \frac{|w|^2}{2 (\KK_1)_{\eta\etabar}} }~.
\end{align}
In particular one can check that $(\KK_1)_{\eta\etabar} = \gamma_{0v}'$.  This $\KK^{\rm reg}$ is consistent with a $\KK$ that solves the sourced Monge--Ampere equation by the same arguments as in section \ref{singularcase}, provided we identify $c = \tilde{R}_7$ as there.  This demonstrates that $(y,w) = (y_-,0)$, $v$ arbitrary, may be identified with the locus of an NS5-brane.  A similar analysis in the $\UU_+$ and $\UU_Z$ patches shows that $\KK_{+,Z}^{\rm reg}$ takes the same form as \eqref{KregNSbrane} as we approach the $NS$ locus, but with $y_- \to y_+$.  Hence $(y,v) = (y_+,0)$, $w$ arbitrary, also corresponds to an NS5-brane source.  This completes the demonstration that the separated NS5-brane geometry is T-dual to the resolved conifold.

There are two interesting pieces of information we have gleaned from this analysis.  First, the precise map between the separation of the NS5-branes, $\Delta y$, and the parameter $a$ of the small resolution is
\begin{equation}\label{resmodulimap}
\Delta y = \frac{2 a^2}{\tilde{R}_7}~.
\end{equation}
Second, this is an excellent example of where we need the modification of the near-brane analysis in \cite{Lunin:2008tf}: the induced metric on the $NS$-brane is $ds^2 = 2 \gamma_{0v}'(|v|^2) dvd\vbar$, which is not flat. This is natural.  As we move further out in $|v|$ we move further away from the $NS'$-brane and the change in gravitational strength is encoded in the warping of the metric.  This is to be contrasted with the singular conifold in which one could find a tangential coordinate $\eta$ such that the induced metric is flat. This was so because the conifold metric is that of a cone.

%%%%%%%%%%%%%%%%%%%%%%%%%%%%
\subsection{The brane web geometry}
%%%%%%%%%%%%%%%%%%%%%%%%%%%%

Let us now present the geometry produced by the pair of NS5-branes described above.  It is determined by the function $\KK$ as dictated in \eqref{NSfive}, \eqref{B2}.  The geometry is singular on the brane locus, as the NS5-branes are fundamental sources for the metric, dilaton, and NS-flux.  On the complement of the locus we have $\KK_{a\bbar} = \KK_{a\bbar}^{\rm reg}$, and in particular $e^{-3A} = \det{(\KK_{a\bbar}^{\rm reg})}$.  Therefore $\KK^{\rm reg}$ as given in \eqref{eq:resKdef} is sufficient for describing the geometry on the complement.

In order to write the geometry explicitly we first change coordinates from the natural brane web coordinate system, $(y,x^7; v,w,\vbar,\wbar)$ to a convenient radial-angular one, $(\rho,\theta_1,\theta_2,\phi_v,\phi_w, \phi_7)$.  The change of coordinates map is
\begin{align}\label{rescov}
& y = \frac{R_7}{6 \ell_{s}^2} \left[ \rho^2(\cos{\theta_1} + \cos{\theta_2}) + 6 a^2 \cos{\theta_2} \right]~, \qquad x^7 = R_7 \phi_7~, \cr
& v = \frac{2 \sqrt{R_7}}{3 \ell_{s}} \rho (\rho^2 + 9 a^2)^{1/4} \cos{\frac{\theta_1}{2}} \sin{\frac{\theta_2}{2}} e^{i \phi_v}~, \cr
& w = \frac{2 \sqrt{R_7}}{3 \ell_{s}} \rho (\rho^2 + 9 a^2)^{1/4} \sin{\frac{\theta_1}{2}} \cos{\frac{\theta_2}{2}} e^{i \phi_w}~.
\end{align}
The coordinates $(\rho,\theta_1,\theta_2)$ are the same ones used to model the resolved conifold via \eqref{eq:TPZmet}, while the phases here are related to the phases appearing there according to $\phi_v = \half (\psi + \phi_1 - \phi_2) - \pi$ and $\phi_w = \half(\psi - \phi_1 + \phi_2)$.  The coordinate $\phi_7$ labels the globally well-defined $U(1)$ direction along which the brane web is smeared; under T-duality it maps to $\phi_x, \phi_Z, -\phi_Y$, or $-\phi_u$ depending on the patch.  The expression for $y$ is consistent with all of the expressions \eqref{eq:yUp}, \eqref{eq:yUm}, \eqref{eq:yxsevenZY} above, after plugging the appropriate coordinate change analogous to \eqref{eq:TPZcov} and making use of $4L/\sqrt{3} = \tilde{R}_7 = \ell_{s}^2/R_7$.  Making the change of variables \eqref{rescov} allows us to bypass the difficulties\footnote{Solving \eqref{eq:yUp} for $|x|^2$ as a function of $(y,|v|,|w|)$ would be equivalent to finding the explicit inverse coordinate map to \eqref{rescov}.} associated with inverting expressions such as \eqref{eq:yUp} to determine $|x|^2$ in terms of $(y,|v|,|w|)$.  As $a \to 0$ we recover the parameterisation used to describe the brane web dual of the singular conifold in \cite{McOrist:2011in}.

The $NS$-brane and $NS'$-brane are located at $(\theta_1,\theta_2) = (0,\pi)$ and $(\pi,0)$ respectively, where we have
\begin{align}\label{resbranelocus}
NS~: \quad & (\theta_1,\theta_2) = (0,\pi)~: \qquad w = 0~, \quad y = -\frac{a^2 R_7}{\ell_{s}^2} = y_- ~, \cr
NS' ~: \quad & (\theta_1,\theta_2) = (\pi,0)~: \qquad v = 0, \qquad y = \frac{a^2 R_7}{\ell_{s}^2} = y_+ ~.
\end{align}
The $\P^1$ sitting at $\rho = 0$ in the resolved conifold is mapped to the segment $y \in [ y_-, y_+ ]$ on the $y$-axis and the $\phi_7$ circle fibration over it:
\begin{equation}\label{P1image}
\rho = 0~: \qquad v = w = 0~, \quad y = \frac{a^2 R_7}{\ell_{s}^2} \cos{\theta_2}~.
\end{equation}
In the resolved conifold the $\phi_Z(\phi_Y)$ circle (or equivalently the $\phi_2$ circle--see \eg\ \eqref{eq:TPZcov}) shrinks to zero at y = $y_+(y_-)$ such that the resulting space is a round $\P^1$.  Here these points are part of the brane locus, and we will see that the $\phi_7$ circle blows up as we approach them.

The dilaton and metric produced by the brane web take the form
\begin{align}\label{resdilaton}
& e^{2(\phi - \phi_0)} = \frac{18 \ell_{s}^4}{R_{7}^2 \ \kappa(\rho) \ \hat{F}(\rho,\theta_1,\theta_2)}~,  \\ \label{resbwmet}
& ds_{6}^2 = \kappa^{-1}(\rho) d\rho^2 + \rho^2 d\hat{\Omega}_{4}^2 + a^2 d\theta_{2}^2 + e^{2(\varphi - \varphi_0)} R_{7}^2 d\phi_{7}^2~,
\end{align}
where $\kappa(\rho)$ is given by \eqref{eq:kappadef},
\begin{equation}\label{Fhatdef}
\hat{F} = \rho^2 \left[ 6(1+c_1 c_2) - (c_1 + c_2)^2\right] + \frac{9 a^2 \rho^2}{\rho^2+9a^2} (s_{2}^2 - s_{1}^2) + \frac{108 a^4}{\rho^2+9a^2} s_{2}^2~,
\end{equation}
and
\begin{align}\label{dOmega4res}
d\hat{\Omega}_{4}^2 =&~ \frac{1}{6}(d\theta_{1}^2 + d\theta_{2}^2) + \frac{\rho^2}{3 \hat{F}}\left( \frac{\rho^2 + 12 a^2}{\rho^2 + 9a^2}\right) s_{1}^2 s_{2}^2 d\phi_v d\phi_w + \cr
& + \frac{2}{3 \hat{F}} \displaystyle\biggl\{ \rho^2 \left[ 7 + c_1 c_2 - 3(c_1 -c_2)\right]  + \cr
& \qquad \quad  +\frac{ 3 a^2}{\rho^2+9a^2} \left[ \rho^2(7+c_1) + 72 a^2 \right](1+c_2) \displaystyle\biggr\} \cos^2{\frac{\theta_1}{2}} \sin^2{\frac{\theta_1}{2}} d\phi_{v}^2 + \cr
&  + \frac{2}{3 \hat{F}} \displaystyle\biggl\{ \rho^2 \left[ 7 + c_1 c_2 + 3(c_1 -c_2)\right]  + \cr
& \qquad \quad  +\frac{ 3 a^2}{\rho^2+9a^2} \left[ \rho^2(7-c_1) + 72 a^2 \right](1-c_2) \displaystyle\biggr\} \sin^2{\frac{\theta_1}{2}} \cos^2{\frac{\theta_1}{2}} d\phi_{w}^2~, \cr
\end{align}
with $c_i \equiv \cos{\theta_i}$ and $s_i \equiv \sin{\theta_i}$.  In the $a\to 0$ limit the separation in $y$ between the NS5-branes goes to zero, and the metric and dilaton reduce directly to those obtained in \cite{McOrist:2011in} for the intersecting NS5-brane system dual to the singular conifold.  As $\rho$ goes to zero, we find that the dilaton $e^{2(\varphi - \varphi_0)} \to \ell_{s}^4/(R_{7}^2 a^2 s_{2}^2)$, and the metric takes the form
\begin{equation}\label{metRescore}
\lim_{\rho \to 0} ds_{6}^2 =a^2 d\theta_{2}^2 + \left( \frac{\ell_{s}^2}{a \sin{\theta_2}}\right)^2 d\phi_{7}^2 + \OO(\rho^2)~.
\end{equation}
This is precisely what one expects from T-duality.  We take the round metric on a $\P^1$ and identify the azimuthal circle fibre with the T-duality circle.  We then invert the radius of the circle fibre, sending the radius to $\ell_{s}^2$ over the radius.

The Neveu--Schwarz $B$-field is patch dependent.  Three patches are required to cover the complement of the brane locus.  We may take these patches to be the images of $\UU_{\pm}$ and $\UU_Z$ under the T-duality map, while the image of $\UU_Y$ is identical to that of $\UU_Z$.  In terms of the brane web coordinate system, $\UU_+$ $(\UU_-)$ covers everything on the complement except points on the $y$-axis satisfying $y < a^2/c$ $(y > -a^2/c)$, while $\UU_Z$ covers everything on the complement except points on the $y$-axis satisfying $|y| > a^2/c$.

On $\UU_+$ the $B$-field is given by
\begin{align}\label{Bresp}
& B_{2}^+ = -\frac{2 \ell_{s}^2}{\hat{F}} \displaystyle\biggl\{ \left[ \left( 3 \kappa(\rho)^{-1} + \frac{\rho^2 c_2}{\rho^2 + 9a^2} -2 c_1 \right) \rho^2 + 18 \kappa(\rho)^{-1} a^2 (1+c_2)  \right]  \sin^2{\frac{\theta_2}{2}} d\phi_v  + \cr
&~ \qquad \qquad  \quad  + \left[ 3 \kappa(\rho)^{-1} + \frac{\rho^2 c_1}{\rho^2 + 9a^2} -2 c_2 \right] \rho^2 \sin^2{\frac{\theta_1}{2}} d\phi_w  \displaystyle\biggr\} \wedge d\phi_7~,
\end{align}
while it differs on other patches by a gauge transformation:
\begin{equation}\label{BresZm}
B_{2}^Z = B_{2}^+ + \ell_{s}^2 d\phi_v \wedge d\phi_7~, \qquad B_{2}^{-} = B_{2}^+ + \ell_{s}^2 d\phi_w \wedge d\phi_7~.
\end{equation}

The gauge transformations encode the nonzero H-charge of the system and can be used to compute it.  Consider a three-cycle enclosing the $NS'$-brane of the form $S_{x^7}^1 \times \Sigma_2$, where $\Sigma_2$ is a two-sphere centered on $(y,v) = (y_+, 0)$, with radius less than $\Delta y$.  On the northern hemisphere the pullback of $B_{2}^+$ is well defined, while on the southern hemisphere the pullback of $B_{2}^Z$ is well defined.  Letting $\d H^+$ denote the boundary of the northern hemisphere, we have
\begin{equation}\label{NSpH3charge}
Q_0 = -\int_{\Sigma_3} H_3 =  -\int_{\d H^+ \times S_{x^7}^1} (B_{2}^+ - B_{2}^Z) = (2\pi \ell_s)^2~,
\end{equation}
which is the NS-charge of a single NS5-brane.  We can also consider a two-sphere centered on $(y,w) = (y_-,0)$ and use the difference $B_{2}^Z - B_{2}^-$ to find the same charge for the $NS$-brane.

%%%%%%%%%%%%%%%%%%%%%%%%%%%%
%%%%%%%%%%%%%%%%%%%%%%%%%%%%
\section{T-dualising the deformed conifold}
%%%%%%%%%%%%%%%%%%%%%%%%%%%%
%%%%%%%%%%%%%%%%%%%%%%%%%%%%

In this section we repeat the story for the deformed conifold and its dual NS5-brane configuration, referred to in the literature as the `diamond web'.  This terminology was introduced in \cite{Aganagic:1999fe} where it was argued that the deformed conifold is T-dual to a brane web described by $v w = -\half \varepsilon^2$, where $\varepsilon$ is the deformation parameter of the conifold.  This profile gives a smoothing out of the singular web, $v w = 0$.  We start off with a quick review of the deformed conifold geometry, following \cite{Candelas:1989js,Minasian:1999tt}. Following sections 2,3, we determine $\KK^{\rm reg}$ via the Buscher rules and a Legendre transform, extend the solution to the source locus, and finally write down the supergravity solution for the brane web.

%%%%%%%%%%%%%%%%%%%%%%%%%%%%%%
\subsection{A desultory discourse on the deformed conifold}
%%%%%%%%%%%%%%%%%%%%%%%%%%%%%%

The deformed conifold may be viewed as a hypersurface in $\mathbb{C}^4$ given by the solution set to
\begin{equation}\label{defsurface}
\det{\WW} = vw - xu = -\half \varepsilon^2~,
\end{equation}
with $\WW$ as in \eqref{eq:resdef}.  Let us define the radial coordinate
\begin{equation}\label{rdeformed}
r^2 = {\rm tr}{\left( \WW^{\dag} \WW\right)} = |x|^2 + |v|^2 + |w|^2 + |u|^2~,
\end{equation}
as usual.  In order to determine the range of $r$, it is helpful to pass to a different coordinate system on $\mathbb{C}^4$.  We define $w^A$, $A = 1,\ldots,4$, such that
\begin{equation}\label{WwA}
\WW = \frac{1}{\sqrt{2}} \left( \begin{array}{c c} w^3 + i w^4 & w^1 - i w^2 \\ w^1 + i w^2 & -w^3 + i w^4 \end{array}\right)~.
\end{equation}
Then, \eqref{defsurface} and \eqref{rdeformed} read $\sum_A (w^A)^2 = \varepsilon^2$ and $\sum_A |w^A|^2 = r^2$.  Letting $x^A = {\rm Re}(w^A)$ and $y^A = {\rm Im}(w^A)$, we get
\begin{equation}\label{wAxy}
\begin{array}{l} x \cdot x - y \cdot y = {\rm Re}(\varepsilon^2) \\ 2 x \cdot y = {\rm Im}(\varepsilon^2) \end{array}~, \qquad x\cdot x + y \cdot y = r^2~,
\end{equation}
from which one may deduce that $r \geq |\varepsilon|$.

Let us now demonstrate that, as in the case of the singular conifold, surfaces of constant $r$, with $r > |\varepsilon|$, admit a transitive $SU(2) \times SU(2)$ action with a $U(1)$ stabiliser.  A particular solution to \eqref{defsurface}, \eqref{rdeformed} is
\begin{equation}\label{Weps}
\WW_{\varepsilon} = \left( \begin{array}{c c} 0 & \half \varepsilon^2 \alpha^{-1} \\ \alpha & 0 \end{array}\right)~, \qquad \textrm{with} \quad \alpha \equiv \half ( \sqrt{ r^2 + |\varepsilon|^2} - \sqrt{r^2 - |\varepsilon|^2} )~.
\end{equation}
The most general solution is
\begin{equation}\label{Wgen}
\WW = L \WW_\varepsilon R^{\dag}~,
\end{equation}
with $L,R \in SU(2)$.  This shows that $SU(2) \times SU(2)$ acts transitively on surfaces of fixed radius. However, certain matrices $(L,R)$ leave $\WW_{\varepsilon}$ fixed. When $r > |\varepsilon|$ these are of the form $(L,R) = (\Theta, \Theta^\dag)$, with $\Theta = {\rm diag}(e^{i\theta}, e^{-i\theta})$.  Hence the set of solutions to \eqref{defsurface} at fixed radius $r > |\varepsilon|$ can be identified with the set of matrices $(L,R) \in SU(2) \times SU(2)$ modulo the equivalence relation $(L,R) \sim (L \Theta, R \Theta^\dag)$.  In particular these surfaces have the topology of $\frac{SU(2) \times SU(2)}{U(1)} \simeq S^2 \times S^3$, just as in the case of the singular conifold. When $r = |\varepsilon|$ on the other hand, we have that
\begin{equation}\label{Wcore}
\WW_{\vareps} = \frac{\vareps}{\sqrt{2}} \left( \begin{array}{c c} 0 & \vareps/|\vareps| \\ \varepsbar/|\vareps| & 0 \end{array}\right) \equiv \frac{\vareps}{\sqrt{2}} \sigma_{\vareps}~,
\end{equation}
with $\sigma_{\vareps} \in SU(2)$.  In this case the stabiliser is an entire $SU(2)$, since if $(L,R) = (L,\sigma_{\vareps}^{\dag} L \sigma_{\vareps})$ then $L \sigma_{\vareps} R^{\dag} = \sigma_{\vareps}$.  Thus, the surface $r = |\varepsilon|$ is an $SU(2) = S^3$.  Heuristically speaking, we have taken the singular conifold and replaced the singular point by an $S^3$.  One way to make this statement more precise is to give the deformed conifold a metric.

In fact, the deformed conifold can be given an explicit Ricci-flat K\"ahler metric, as first demonstrated in \cite{Candelas:1989js}.  As in the case of the singular conifold, requiring the K\"ahler potential to be invariant under the $SU(2) \times SU(2)$ action implies that it should only depend on the radial coordinate $r^2$ as defined in  \eqref{rdeformed}.  We introduce the usual patches, $\UU_+$ where $x \neq 0$, and $\UU_-$ where $u \neq 0$.  On $\UU_+$ we can solve \eqref{defsurface} for $u$ and use $z^{\alpha}= (x,v,w)$ as a complex coordinate system, while on $\UU_-$ we eliminate $x$ and use $z^\alpha = (u,v,w)$ to parameterise the patch.  On the overlap we have a transition map of the appropriate form, \eqref{yKTs}, and the set of points not covered by $\UU_+ \cup \UU_-$ is our proposed source locus: $\{ x = u = v w + \half \varepsilon^2 = 0 \}$.  Denote the K\"ahler potential by $\FF^{\flat} = \FF^{\flat}(r^2; \varepsilon)$.  Then on the upper patch the K\"ahler metric takes the form
\begin{align}\label{defKahmet}
& (ds^{\flat})^2 = 2 g_{\alpha\betabar} dz^\alpha d\zbar^{\betabar} = 2 \left[ (\d_\alpha \delbar_{\betabar} r^2) (\FF_{+}^{\flat})' + (\d_\alpha r^2)(\delbar_{\betabar} r^2) (\FF_{+}^{\flat})'' \right] dz^\alpha d\zbar^{\betabar}~,
\end{align}
where
\begin{equation}\label{rsqUpdef}
r^2 = |x|^2 + |v|^2 + |w|^2 + \frac{ |v w + \half \varepsilon^2 |^2}{|x|^2}~.
\end{equation}
Here the primes denote differentiation with respect to $r^2$.  An identical expression holds on the lower patch with $x \to u$.

Using these expressions one can straightforwardly compute $g = \det{(g_{\alpha\betabar})}$.  The condition for Ricci-flatness, $\d_\alpha \delbar_{\betabar} \log{g} = 0$, yields an equation for $\FF^{\flat}$:
\begin{align}\label{Rflatdef}
& r^2 (r^4- |\varepsilon|^4) (\gamma^{\flat})^{3}{}' + 3 |\varepsilon|^4 (\gamma^{\flat})^3 = 2 L^2 r^8~, \qquad \textrm{where} \quad \gamma^{\flat} = r^2 (\FF^{\flat})'~.
\end{align}
By requiring that $\FF^{\flat}(r^2;0) = \FF^{\sharp}$, we can identify $L$ with the same integration constant introduced in the case of the singular conifold.  After changing the radial variable to $\tau \in [0, \infty)$,
\begin{equation}\label{taudef}
r^2 = |\varepsilon|^2 \cosh{\tau}~,
\end{equation}
one finds that \eqref{Rflatdef} can be integrated, yielding
\begin{align}\label{gammadef}
& \gamma^{\flat} = \left( \frac{L^2 |\varepsilon|^4}{2} \right)^{1/3} \frac{ (\sinh{(2\tau)} - 2\tau )^{1/3}}{\tanh{\tau}} \cr
\Rightarrow \quad & \FF^{\flat} = \left( \frac{L^2 |\varepsilon|^4}{2} \right)^{1/3} \int_{0}^{\tau} d\tilde{\tau} \left( \sinh{(2\tilde{\tau})} - 2 \tilde{\tau} \right)^{1/3}~.
\end{align}
While the final integral is not known, the analytic form of $\gamma^{\flat}$ is sufficient for writing down the metric explicitly.

In order to put the metric in a manageable form, we introduce Euler angles such that $L = e^{-i \phi_1 \sigma^3/2} e^{i \theta_1 \sigma^2/2} e^{-i \psi_1 \sigma^3/2}$ and $R = e^{-i \phi_2 \sigma^3/2} e^{i \theta_2 \sigma^2/2} e^{-i \psi_2 \sigma^3/2}$.  Plugging into \eqref{Wgen}, and noting that $\alpha = |\vareps| e^{-\tau/2}/\sqrt{2}$, we have the following change of variables:
\begin{align}\label{defcov}
& x = \frac{\varepsilon}{\sqrt{2}} \left(  \cos{\frac{\theta_1}{2}} \cos{\frac{\theta_2}{2}} e^{\half (\tau + i \psi)} - \sin{\frac{\theta_1}{2}} \sin{\frac{\theta_2}{2}} e^{-\half (\tau + i \psi)} \right) e^{\frac{i}{2} (\phi_1 + \phi_2)} ~, \cr
& v = -  \frac{\varepsilon}{\sqrt{2}} \left(  \cos{\frac{\theta_1}{2}} \sin{\frac{\theta_2}{2}} e^{\half (\tau + i \psi)} + \sin{\frac{\theta_1}{2}} \cos{\frac{\theta_2}{2}} e^{-\half (\tau + i \psi)} \right) e^{\frac{i}{2} (\phi_1 - \phi_2)}~, \cr
& w = \frac{\varepsilon}{\sqrt{2}} \left(  \sin{\frac{\theta_1}{2}} \cos{\frac{\theta_2}{2}} e^{\half (\tau + i \psi)} + \cos{\frac{\theta_1}{2}} \sin{\frac{\theta_2}{2}} e^{-\half (\tau + i \psi)} \right) e^{\frac{i}{2} ( - \phi_1 + \phi_2)} ~, \cr
& u =  \frac{\varepsilon}{\sqrt{2}} \left( - \sin{\frac{\theta_1}{2}} \sin{\frac{\theta_2}{2}} e^{\half (\tau + i \psi)} + \cos{\frac{\theta_1}{2}} \cos{\frac{\theta_2}{2}} e^{-\half (\tau + i \psi)} \right) e^{-\frac{i}{2} (\phi_1 + \phi_2)}~,
\end{align}
where $\psi \equiv \psi_1 + \psi_2 +2 {\rm Arg}(\vareps)$.  As $\varepsilon \to 0$, $\tau \to \infty$ and we recover the parameterisation for the singular conifold.

Using \eqref{gammadef}-\eqref{defcov}, we find that the metric \eqref{defKahmet} takes the form given in \cite{Minasian:1999tt,Klebanov:2000hb}:
\begin{align}\label{defcon}
(ds^\flat)^2 =&~ (L^2 |\varepsilon|^4)^{1/3} K(\tau) \displaystyle\biggl\{ \frac{1}{3 K(\tau)^3} \left[ d\tau^2 + (g^5)^2 \right] + \sinh^2{\left(\frac{\tau}{2}\right)} \left[ (g^{1})^2 + (g^{2})^2 \right] + \cr
& \qquad \qquad \qquad  \qquad +  \cosh^2{\left(\frac{\tau}{2}\right)} \left[ (g^{3})^2 + (g^{4})^2 \right]  \displaystyle\biggr\}~,
\end{align}
where
\begin{align}\label{Ktau}
& K(\tau) = \frac{ (\sinh{(2\tau)} - 2\tau)^{1/3} }{2^{1/3} \ \sinh{\tau}}~, \\ \label{goneforms}
& g^{1,3} = \frac{e^1 \mp e^3}{\sqrt{2}}~, \qquad g^{2,4} = \frac{e^2 \mp e^4}{\sqrt{2}}~, \qquad g^5 = e^5~,
\end{align}
with
\begin{align}\label{eoneforms}
& e^1 = -\sin{\theta_1} d\phi_1~, \qquad e^2 = d\theta_1~, \cr
& e^3 =  \cos{\psi} \sin{\theta_2} d\phi_2 - \sin{\psi} d\theta_2  ~, \cr
& e^4 = \sin{\psi} \sin{\theta_2} d\phi_2 + \cos{\psi} d\theta_2 ~, \cr
& e^5 = d\psi + \cos{\theta_1} d\phi_1 + \cos{\theta_2} d\phi_2~.
\end{align}
The same expression can be obtained starting from the K\"ahler metric in the lower patch, \eqref{defKahmet} with $x \to u$.

In the limit $r/|\varepsilon| \to \infty$, such that $e^{\tau} \sim r^2/|\varepsilon|^2$, it is easy to show \eqref{defcon} approaches the metric of the singular conifold.  In the limit $r \to |\varepsilon|$, corresponding to $\tau \to 0$, we find that
\begin{equation}\label{defconsmalltau}
(ds^\flat)^2 = \frac{1}{2} \left( \tilde{R}_{7}^2 |\varepsilon|^4 \right)^{1/3} \left( \half (g^{5})^2 + (g^{3})^2 + (g^{4})^2 \right) + \OO(\tau^2)~,
\end{equation}
where $\tilde{R}_7 = 4L/\sqrt{3}$.  The leading order terms correspond to the round metric on a three-sphere of radius $\tilde{R}_{7}^{1/3} |\varepsilon|^{2/3}$.  To see this we observe that \cite{Candelas:1989js,Minasian:1999tt}
\begin{equation}\label{gTrel}
\half (g^5)^2 + (g^3)^2 + (g^4)^2 = {\rm tr} \left( dT^{\dag} dT \right)~,
\end{equation}
where $T \in SU(2)$ is given by
\begin{equation}\label{Tdef}
 T = L \sigma^1 R^\dag \sigma^1 = \lim_{\tau \to 0} \frac{\sqrt{2}}{\varepsilon} \left(\begin{array}{c c} x & v \\ w & u \end{array}\right)~.
\end{equation}

Later it will be useful to have a more explicit parameterisation of the $S^3$ at $\tau = 0$.  We write
\begin{equation}\label{Teuler}
T = \left( \begin{array}{c c} \cos{\frac{\vartheta}{2}} e^{i \tilde{\phi}_x} & -\sin{\frac{\vartheta}{2}} e^{-i \tilde{\phi}_w} \\   \sin{\frac{\vartheta}{2}} e^{i \tilde{\phi}_w} & \cos{\frac{\vartheta}{2}} e^{-i \tilde{\phi}_x} \end{array}\right)~,
\end{equation}
where $\tilde{\phi}_{x,w} = \phi_{x,w} - {\rm Arg}(\varepsilon)$, such that
\begin{equation}\label{dTdTcanonical}
\half {\rm tr}\left( dT^{\dag} dT\right) = \frac{1}{4} d\vartheta^2 + \sin^2{\frac{\vartheta}{2}} d\phi_{w}^2 + \cos^2{\frac{\vartheta}{2}} d\phi_{x}^2~.
\end{equation}
The three-sphere takes the form of a circle fibration over a disk.  The disk is represented by the first two terms while $\phi_x$ is the fibre coordinate.  The circle fibre is finite at the centre of the disk and shrinks to zero at the boundary such that the total space is smooth.  The relationship between these coordinates and the ones parameterising the rest of the conifold is
\begin{align}\label{S3coords}
& \cos^2{\frac{\vartheta}{2}} = \half \left[ 1 + \cos{\theta_1} \cos{\theta_2} - \cos{\psi} \sin{\theta_1} \sin{\theta_2} \right]~, \cr
& \tilde{\phi}_x = \arctan{\left[ \frac{ \cos{\left( \frac{\theta_1 - \theta_2}{2}\right)} }{  \cos{\left( \frac{\theta_1 + \theta_2}{2}\right)} }  \tan{\frac{\psi}{2}} \right]} + \half (\phi_1 + \phi_2)~, \cr
&  \tilde{\phi}_w = \arctan{\left[ \frac{ \sin{\left( \frac{\theta_1 - \theta_2}{2}\right)} }{  \sin{\left( \frac{\theta_1 + \theta_2}{2}\right)} }  \tan{\frac{\psi}{2}} \right]} - \half (\phi_1 - \phi_2)~.
\end{align}
With this change of coordinates one can verify directly the consistency of \eqref{dTdTcanonical} and \eqref{gTrel}.

%%%%%%%%%%%%%%%%%%%%%%%%%%%%%%%%%%%%%%%%%%%
\subsection{Legendre transformations and the source locus}
%%%%%%%%%%%%%%%%%%%%%%%%%%%%%%%%%%%%%%%%%%%

We define the symplectic coordinate pair and regularised brane potential patchwise in the usual fashion:
\begin{align}\label{ydefcon}
\UU_+~: \quad c y =&~ |x|^2 \d_{|x|^2} \FF_{+}^{\flat}  = \frac{|x|^4 - |v w + \half \vareps^2|^2}{|x|^4 + |x|^2(|v|^2 + |w|^2) + |vw + \half \vareps^2|^2} \gamma^\flat(r^2;\vareps)~, \cr
\UU_-~: \quad c y =&~ - |u|^2 \d_{|u|^2} \FF_{-}^{\flat}  = - \frac{|u|^4 - |v w + \half \vareps^2|^2}{|u|^4 + |u|^2(|v|^2 + |w|^2) + |vw + \half \vareps^2|^2} \gamma^\flat(r^2;\vareps)~, \cr
\end{align}
\begin{align}\label{x7defcon}
& \UU_+~: \quad \tilde{x}^{7(+)} = c \phi_x~, \qquad \UU_-~: \quad \tilde{x}^{7(-)} = - c \phi_u~,
\end{align}
and
\begin{align}\label{Kregdef}
\UU_+~: \quad \KK_{+}^{\rm reg} = \left\{ \FF_{+}^{\flat} - c y \log{(|x|^2/c^2)} \right\}_{|x| = |x|(y,|v|,|w|)}~, \cr
\UU_-~: \quad  \KK_{-}^{\rm reg} = \left\{ \FF_{-}^{\flat} + c y \log{(|u|^2/c^2)} \right\}_{|u| = |u|(y,|v|,|w|)}~.
\end{align}
The definitions of $y$ agree on the overlap $\UU_+ \cap \UU_-$, while $\KK$ and $\tilde{x}^7$ transform according to \eqref{yKTs} with $f_1 = c \log{((v w + \half \vareps^2)/c^2)}$.

The proposed source locus is at $y = vw + \half \vareps^2 = 0$.  Let us determine the behaviour of $\KK^{\rm reg}$ near this locus to see if it is consistent with the required singularity structure of $\KK$.  We suppose $y = \OO(\epsilon)$ and $v w + \half \vareps^2 = \OO(\epsilon)$ and work perturbatively in $\epsilon$.  From \eqref{ydefcon} the leading behaviour of $|x|^2,|u|^2$ is $\OO(\epsilon)$ and given by
\begin{equation}\label{xudef}
|X_{\pm}|^2 = \frac{c r_{0}^2}{2 \gamma_{0}^{\flat}} \left[ \pm y + \sqrt{ y^2 + \frac{4 (\gamma_{0}^{\flat})^2}{c^2 r_{0}^4} \left| v w + \half \vareps^2 \right|^2 } \ \right] + \OO(\epsilon^2)~,
\end{equation}
where $X_+ = x$, $X_- = u$,
\begin{equation}\label{r0def}
r_{0}^2 = r^2 \displaystyle\bigg|_{\epsilon = 0} = \left( |v|^2 + |w|^2 \right) \displaystyle\bigg|_{\epsilon = 0}~,
\end{equation}
and $\gamma_{0}^{\flat} \equiv \gamma^{\flat}(r_{0}^2 ; \vareps)$.  A natural ansatz for the near-brane coordinates $(\eta,\lambda)$ is
\begin{equation}\label{nbdefcoord}
\begin{array}{l} \eta = (c/2) \log{(v/w)} \\ \lambda = v w/c \end{array} \quad \Rightarrow \quad \begin{array}{l} v = \sqrt{c \lambda} \ e^{\eta/c} \\ w = \sqrt{c \lambda} \ e^{-\eta/c} \end{array}~,
\end{equation}
with the source locus at $\lambda = \lambda_0 \equiv - \vareps^2/(2c)$.  In terms of these coordinates $$r_{0}^2 = 2c|\lambda_0| \cosh{((\eta+\etabar)/c)} = |\vareps|^2 \cosh{((\eta+\etabar)/c)}.$$
Expanding $\KK^{\rm reg}$ and using \eqref{xudef} we find that it takes the form
\begin{align}\label{Kregdefnb}
\KK_{\pm}^{\rm reg} =&~ \KK_1 + c \left\{ D \mp y \log{\left[ \sqrt{2 (\KK_1)_{\eta\etabar}} (\pm y + D) \right]} \right\} + \cr
&+ y h_1(\eta,\etabar) + 2 {\rm Re}\left[(\lambda - \lambda_0)h_2(\eta,\etabar)\right] + \OO(\epsilon^2)~, \qquad \textrm{where} \cr
D =&~ \sqrt{y^2 + \frac{|\lambda - \lambda_0|^2}{2 (\KK_1)_{\eta\etabar}} }~,
\end{align}
and we have identified $\KK_1 = \FF_{0}^{\flat} \equiv \FF^{\flat}(r_{0}^2;\vareps)$. A nice check of this result is to observe the tangential derivative of $\KK_1$ is consistent with \eqref{xudef}: $(\KK_1)_{\eta\etabar} = r_{0}^4/8 (\gamma_{0}^{\flat})^2$.  From our discussion in section 2, upon identifying $c = \tilde{R}_7$, a $\KK^{\rm reg}$ of the form \eqref{Kregdefnb} is consistent with a $\KK$ corresponding to a charge one brane web with source locus $y = 0, \lambda = \lambda_0$.

This completes the demonstration of the T-duality relation between the deformed conifold and the ``diamond'' brane web at the level of supergravity solutions.  In the next section we will give an explicit parameterisation of the brane web geometry.  We note here that, like the case of the resolved conifold, the induced metric on the brane worldvolume is non-trivial.  Written out explicitly we have
\begin{equation}\label{defindmet}
ds_{\rm ind}^2 = 2 (\KK_1)_{\eta\etabar} d\eta d\etabar = \left( \frac{4 |\vareps|^2}{3 \tilde{R}_{7}^2} \right)^{2/3} \frac{ \sinh^2{(2 \eta_1)} }{ \left[ \sinh{(4\eta_1)} - 4\eta_1 \right]^{2/3}} (d\eta_{1}^2 + d\eta_{2}^2)~,
\end{equation}
where $\eta = \eta_1 + i \eta_2$.  The worldvolume has the topology of a cylinder, with $\eta_2 \sim \eta_2 + 2\pi \tilde{R}_7$ parameterising the circle direction.  The circle at $\eta_1 = 0$ has minimal radius $\tilde{R}_{7}^{1/3} |\vareps|^{2/3}$.  As we will see below, it can be identified with the boundary of the disk in the presentation of the three-sphere at $\tau=0$ as a circle fibration over a disk.  As $\eta_1 \to +\infty(-\infty)$, the $\eta$-plane approaches the $v(w)$-plane, in accordance with \eqref{nbdefcoord}.

%%%%%%%%%%%%%%%%%%%%%%%%
\subsection{The diamond brane web geometry}
%%%%%%%%%%%%%%%%%%%%%%%%

In order to write the supergravity solution in completely explicit form we employ a radial-angular coordinate system based on the one used to model the deformed conifold geometry.  The change of coordinates from the natural brane-web coordinate system, $(y,x^7;v,w,\vbar,\wbar)$, to the radial-angular one, $(\tau,\theta_1,\theta_2,\psi,\phi_-, \phi_7)$, is
\begin{align}\label{defbwcoords}
& y = \frac{1}{4} \left( \frac{3 R_7 |\vareps|^4}{4 \ell_{s}^2} \right)^{1/3} \left(\sinh{(2\tau)} -2\tau \right)^{1/3} (\cos{\theta_1} + \cos{\theta_2} )~, \qquad x^7 = R_7 \phi_7~, \cr
& v = -  \frac{\varepsilon}{\sqrt{2}} \left(  \cos{\frac{\theta_1}{2}} \sin{\frac{\theta_2}{2}} e^{\half (\tau + i \psi)} + \sin{\frac{\theta_1}{2}} \cos{\frac{\theta_2}{2}} e^{-\half (\tau + i \psi)} \right) e^{i \phi_-/2}~, \cr
& w = \frac{\varepsilon}{\sqrt{2}} \left(  \sin{\frac{\theta_1}{2}} \cos{\frac{\theta_2}{2}} e^{\half (\tau + i \psi)} + \cos{\frac{\theta_1}{2}} \sin{\frac{\theta_2}{2}} e^{-\half (\tau + i \psi)} \right) e^{-i \phi_-/2} ~.
\end{align}
The coordinates $(\tau,\theta_1,\theta_2,\psi)$ are the same as those used in the parameterisation of the deformed conifold.  In particular the expression for $y$ is consistent with \eqref{ydefcon}, after plugging in \eqref{defcov} and using $c = 4L/\sqrt{3} = \ell_{s}^2/R_7$.  We have exchanged the circle coordinates $(\phi_1,\phi_2)$ in favour of $\phi_- \equiv \phi_1 - \phi_2$ and $\phi_7$.  The latter is the globally defined $U(1)$ coordinate along which the brane web is smeared.  These are more convenient as they parameterise the $U(1)$ isometry directions of the brane web: $\phi_7$ labels the T-duality circle, (under T-duality it maps to $\phi_x$ in $\UU_+$ and $-\phi_u$ in $\UU_-$), while $\phi_-$ labels the $U(1)$ generated by $v \to \lambda v$, $w \to \lambda^{-1} w$.

The brane locus is located at $y = \lambda - \lambda_0 = 0$, where recall $c (\lambda - \lambda_0) = v w + \half \vareps^2$.  We have that
\begin{equation}\label{lambdarad}
v w + \half \vareps^2 = \frac{\varepsilon^{2}}{2} \left[ (1+ c_1 c_2 - s_1 s_2\cosh{\tau} \cos{\psi}) - i  s_1 s_2 \sinh{\tau} \sin{\psi} \right]~,
\end{equation}
with $c_i \equiv \cos{\theta_i}$, $s_i \equiv \sin{\theta_i}$.  Given this expression and the one for $y$, we find that the brane locus corresponds to $(\theta_1, \theta_2) = (0,\pi)$ or $(\pi,0)$.  On the locus we have the following expressions for $v,w$:
\begin{align}\label{branelocus}
(\theta_1, \theta_2) = (0,\pi)~: \qquad v = -\frac{\vareps}{\sqrt{2}} e^{\tau/2} e^{\frac{i}{2}(\phi_- + \psi)}~, \quad w = \frac{\vareps}{\sqrt{2}} e^{-\tau/2} e^{-\frac{i}{2}(\phi_- + \psi)}~, \cr
(\theta_1, \theta_2) = (\pi,0)~: \qquad v = -\frac{\vareps}{\sqrt{2}} e^{-\tau/2} e^{\frac{i}{2}( \phi_- - \psi)}~, \quad w = \frac{\vareps}{\sqrt{2}} e^{\tau/2} e^{-\frac{i}{2}(\phi_- - \psi)}~.
\end{align}
In terms of the worldvolume coordinate $\eta$, the first component maps to the upper half cylinder, $\eta_1 \geq 0$, and the second to the lower half, $\eta_1 \leq 0$.  We have $|\eta_1| = c \tau/2$ and $\eta_2 = c (\phi_- \pm \psi)/2$, with the sign depending on the sign of $\eta_1$.  It is clear that $\phi_-$ generates the $U(1)$ isometry of the web.  In order to understand the role of $\psi$, consider \eqref{lambdarad} and specialise to the $y=0$ plane where $\theta_1 + \theta_2 = \pi$:
\begin{align}\label{y0lambda}
y=0 ~(\theta_+ = \pi)~: \quad v w + \half \vareps^2 = \frac{ \vareps^2 }{2} \cos^2{\frac{\theta_-}{2}}  \left[ 1 - \cosh{\tau} \cos{\psi} - i \sinh{\tau} \sin{\psi} \right]~,
\end{align}
where $\theta_{\pm} \equiv \theta_1 \pm \theta_2$.  As $\psi \to \psi + 2\pi$ we make a closed loop in the complex $\lambda$-plane enclosing the point $\lambda_0$; in other words we encircle the brane web.

\begin{figure}
 \includegraphics[bb=0 0 300 280]{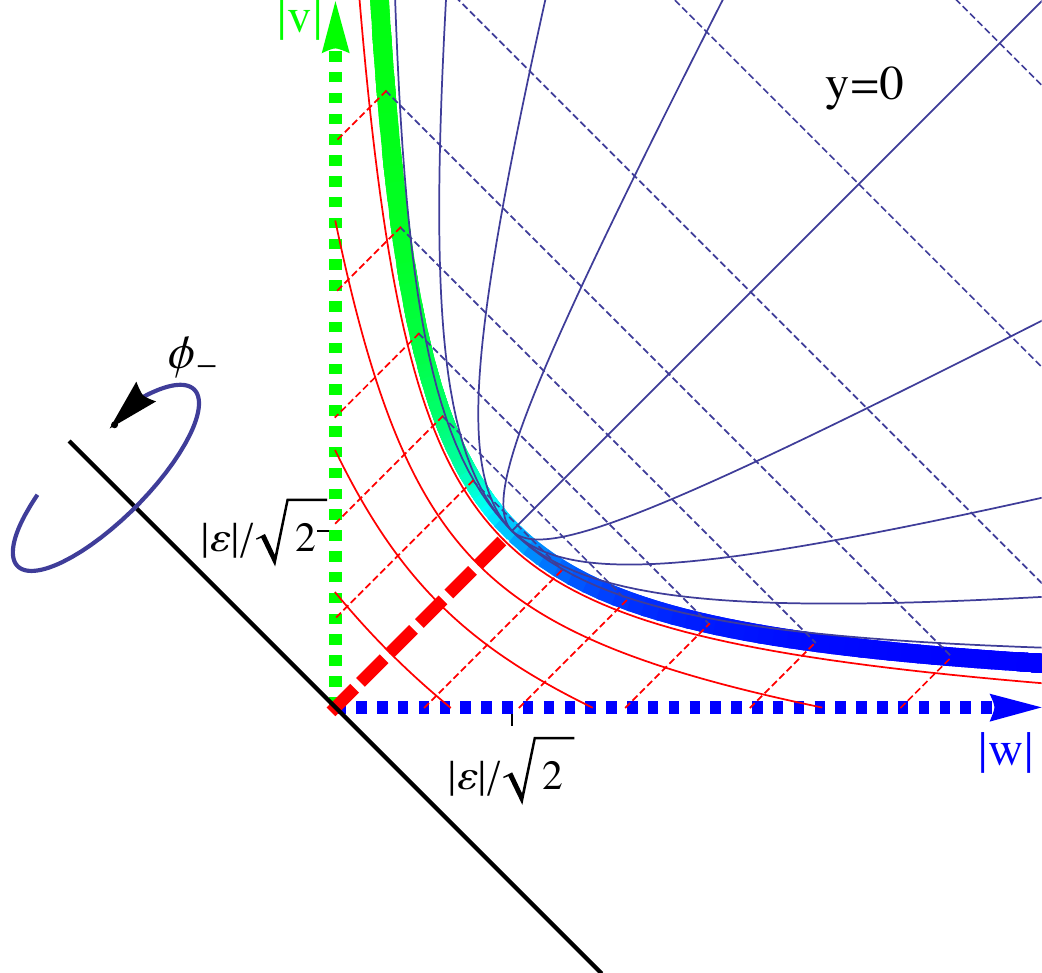}
 \caption{The deformed brane web geometry in the $y=0$ plane.  The brane locus is represented by the solid blue-green curve.  Far from the origin it approaches the $v$-plane (green) or the $w$-plane (blue).  Near the origin the shape is deformed to $|v| |w| = \half |\vareps|^2$.  We can visualise the cylindrical topology of the web by rotating the entire figure about the axis indicated by $\phi_-$.  The rotation generates a disk from the thick dashed (red) line segment on the diagonal.  This disk, together with the T-duality circle fibred over it, is the image (under the T-duality map) of the three-sphere at the core of the deformed conifold.  The grid lines indicate surfaces of constant $\theta_-$ (solid lines) and surfaces of constant $\tau$ (dotted lines).  The blue grid lines to the right of the locus are evaluated at $\psi = 0$ and the red grid lines to the left are evaluated at $\psi = \pi$.  Thus in accordance with \eqref{y0lambda} we are taking a real slice of the near-brane coordinate $(\lambda - \lambda_0)/\vareps^2$.  If we imagine the brane locus as a current carrying wire, $\psi$-circles may be visualised as magnetic field lines encircling the wire.}
 \label{figure1}
 \end{figure}

It is also interesting to understand how the $S^3$ at the core of the deformed conifold maps under T-duality to the brane web configuration.  This $S^3$ sits at $\tau = 0$ and is thus mapped into the $y = 0$ plane.  Furthermore, when $\tau = 0$ it is easy to check that $v,w$ have the same magnitude and opposite phase, up to a shift by $\pi$.  Specifically,
\begin{equation}\label{tauzero}
\tau = 0~: \qquad y = 0~, \quad v = - \frac{|\varepsilon|}{\sqrt{2}} \sin{\frac{\vartheta}{2}} e^{-i \phi_w}~, \quad w = \frac{|\vareps|}{\sqrt{2}}  \sin{\frac{\vartheta}{2}} e^{i \phi_w}~,
\end{equation}
where $(\vartheta,\phi_w)$ are given in terms of $(\theta_i, \psi, \phi_-)$ in \eqref{S3coords}.  Recall that if we view the three-sphere as a circle fibration over a disk, \eqref{dTdTcanonical}, $\phi_w$ is the plane-polar angle of the disk while $\phi_x$, which we can identify with $\phi_7$ under the T-duality map, is the fibre coordinate.  We see that the centre of the disk maps to the origin of brane web coordinates, $y = v= w = 0$.  Meanwhile the boundary of the disk, where $|v| = |w| = |\vareps|/\sqrt{2}$, maps onto the brane locus.  Specifically it maps onto the minimal size circle of the brane locus at $\tau = 0$.  On the deformed conifold side of the T-duality, the $\phi_x$ circle shrinks to zero at the boundary of the disk such that we have a round $S^3$.  On the brane web side, however, we expect it to blow up at the source locus.  We will see explicitly that this is the case when we write the metric below.  Figure \ref{figure1} gives a representation of the geometry in the $y=0$ plane.

Let us now present the supergravity background produced by the web.  The dilaton, metric, and NS-NS flux are all determined by the brane potential $\KK$, according to \eqref{NSfive}.  More precisely, they all depend on derivatives of type $\KK_{a\bbar}$.  On the complement of the brane locus, $\KK_{a\bbar} = \KK_{a\bbar}^{\rm reg}$, so \eqref{Kregdef} is sufficient for determining the geometry.

We begin with the dilaton:
\begin{equation}\label{defdilaton}
e^{2(\varphi - \varphi_0)} = 2\cdot 12^{2/3} \left( \frac{\ell_{s}^2}{R_7 |\varepsilon|} \right)^{4/3} \frac{ K(\tau)^2}{F^{\flat}(\tau,\theta_1,\theta_2,\psi)}~,
\end{equation}
where $K(\tau)$ is given in \eqref{Ktau} and
\begin{equation}\label{Fdef}
F^{\flat}(\tau,\theta_1,\theta_2,\psi) \equiv 2 (c_1+c_2)^2 + 3 K(\tau)^3 \left[ ch_{\tau} \left( s_{1}^2 + s_{2}^2 \right) - 2 c_\psi  s_1 s_2 \right]~.
\end{equation}
Here we have introduced the additional shorthand $c_{\psi} \equiv \cos_{\psi}$, $ch_{\tau} = \cosh{\tau}$, and similarly for the (hyperbolic) sine functions.  The function $F^{\flat}$, which also appears in the metric, is zero along the brane locus at $(\theta_1,\theta_2) = (0,\pi)$ and $(\pi,0)$ and otherwise positive.  Consequently the dilaton blows up as we approach the brane web, but is otherwise finite.  $F^{\flat}$ has the following behaviour for small and large $\tau$:
\begin{align}\label{Flimits}
& \lim_{\tau \to 0} F^{\flat} = 8 \cos^2{\frac{\vartheta}{2}}~, \cr
& \lim_{\tau \to \infty} F^{\flat} = 6 (1 + c_1 c_2) - (c_1 + c_2)^2~,
\end{align}
where again $\vartheta$ is the $S^3$ coordinate defined in \eqref{S3coords}.  The corresponding behaviour of the dilaton is
\begin{align}\label{defdillim}
& \lim_{\tau \to 0} e^{2(\varphi - \varphi_0)} = \left( \frac{\ell_{s}^2}{R_7 |\vareps| } \right)^{4/3} \sec^2{\frac{\vartheta}{2}}~, \cr
& \lim_{\tau \to \infty} e^{2(\varphi - \varphi_0)} = \frac{18 \ell_{s}^4}{R_{7}^2 \rho^2 \left[ 6(1+ c_1 c_2) - (c_1 + c_2)^2\right] }~,
\end{align}
where $\rho \equiv \sqrt{3} L^{1/3} r^{2/3} \to \frac{3}{2} \left( \frac{\tilde{R}_7 |\vareps|^2}{3} \right)^{1/3} e^{\tau/3}$ is the radial coordinate used for the singular and resolved conifold.  Thus for $r \gg |\vareps|$, the dilaton matches onto the form we obtained for the brane web dual to the singular conifold \cite{McOrist:2011in}.  Meanwhile, on the (image of the) $S^3$ at $r = |\vareps|$, it blows up at the boundary of the disk where $S^3$ and brane locus intersect.

The metric has a structure familiar from the singular and resolved conifold,
\begin{equation}\label{defbwmet}
ds_{6}^2 = \left( L^2 |\vareps|^4 \right)^{1/3} K(\tau) \left[ \frac{1}{3 K(\tau)^3} d\tau^2 + d\Omega_{4}^{\flat 2} \right] + e^{2(\varphi - \varphi_0)} R_{7}^2 d\phi_{7}^2~,
\end{equation}
with the compact four-manifold given by
\begin{align}\label{Omega4def}
& d\Omega_{4}^{\flat 2} = \half ch_{\tau} \left(d\theta_{1}^2 + d\theta_{2}^2 \right) + c_{\psi} d\theta_1 d\theta_2  - \frac{3 K(\tau)^3}{2 F^{\flat}} s_{\psi}^2 (s_2 d\theta_1 + s_1 d\theta_2)^2 +  \cr
& ~ -\frac{2 s_{\psi} (c_1 + c_2)}{F} \left[ (s_2 d\theta_1 + s_1 d\theta_2) d\psi + (c_1 s_2 d\theta_1 - c_2 s_1 d\theta_2) d\phi_- \right] + \cr
& ~  - \frac{3 K(\tau)^3}{F^{\flat}} s_{\psi} s_1 s_2 \left[ ch_{\tau} \left( s_1 d\theta_1 - s_2 d\theta_2 \right) - c_{\psi}  (s_2 d\theta_1 - s_1 d\theta_2) \right] d\phi_- + \cr
& ~  + \frac{1}{F^{\flat}} \displaystyle\biggl\{ \left[ ch_{\tau} (s_{1}^2 + s_{2}^2) - 2 c_{\psi} s_1 s_2 \right] d\psi^2 + 2 \left[ ch_{\tau} (1+c_1c_2) - c_{\psi} s_1 s_2 \right] (c_1 - c_2) d\psi d\phi_- + \cr
& \qquad ~~ + \left[ ch_{\tau} ( s_{1}^2 c_{2}^2 + s_{2}^2 c_{1}^2 )  + 2 c_{\psi} s_1 c_1 s_2 c_2 + \frac{3}{2} K(\tau)^3 \left( sh_{\tau}^2 + s_{\psi}^2 \right) s_{1}^2 s_{2}^2 \right] d\phi_{-}^2 \displaystyle\biggr\}~. \end{align}
We again have an torus fibration over the $\theta_1$-$\theta_2$ base, $[0,\pi] \times [0,\pi]$.  The detailed form of the metric is somewhat more complicated due to the lower amount of symmetry preserved by the web.  The fibration becomes singular along the brane locus at $(\theta_1,\theta_2) = (0,\pi)$ and $(\pi,0)$, and is smooth everywhere else.

In the $\tau \to \infty$ limit of \eqref{defbwmet} we recover the metric for the brane web dual to the singular conifold.  For large $\tau$, the terms that dominate \eqref{Omega4def} are those going as $e^{\tau}$ where note that $K(\tau) \to 2^{1/3} e^{-\tau/3}$.  We have
\begin{align}\label{Omega4lt}
d\Omega_{4}^{\flat 2} =& \displaystyle\biggl\{ \frac{1}{4} \left( d\theta_{1}^2 + d\theta_{2}^2 \right) + \frac{1}{2 F^{\sharp}} \displaystyle\bigg[ (s_{1}^2 + s_{2}^2) d\psi^2 + 2(1+c_1 c_2)(c_1 - c_2) d\psi d\phi_- + \cr
& \qquad \qquad \qquad \qquad \qquad + \left( s_{1}^2 c_{2}^2 + s_{2}^2 c_{1}^2 + \frac{3}{2} s_{1}^2 s_{2}^2 \right) d\phi_{-}^2 \displaystyle\biggr]  \displaystyle\biggr\} e^{\tau} +  \OO(1)~,
\end{align}
where $F^{\sharp} = 6(1+c_1 c_2) - (c_1 + c_2)^2$.  As $\tau \to \infty$ we learn from \eqref{defbwcoords} that the phases of $v$ and $w$ satisfy
\begin{equation}\label{vwphaseslt}
d\phi_v = \half (d\psi + d\phi_-) + \OO(e^{-\tau})~, \qquad d\phi_w = \half (d\psi - d\phi_-) + \OO(e^{-\tau})~.
\end{equation}
Taking this into account and using the definition of $\rho$ as in \eqref{defdillim} we eventually find that the metric can be put in a form consistent with the brane web dual of the singular conifold \cite{McOrist:2011in}.  Notice the emergence of the second $U(1)$ isometry in $d\Omega_{4}^{\flat 2}$ as $\tau \to \infty$.

In the $r \to |\vareps|$, or $\tau \to 0$ limit, it can be shown that $d\Omega_{4}^{\flat 2}$ collapses to the metric on the two-dimensional disk:
\begin{equation}\label{Omega4st}
d\Omega_{4}^{\flat 2} = 2 \left( \frac{1}{4} d\vartheta^2 + \sin^{2}{\frac{\vartheta}{2}} d\phi_{w}^2 \right) + \OO(\tau^2)~,
\end{equation}
where the expressions for $\vartheta,\phi_w$ in terms of $(\theta_1,\theta_2,\psi,\phi_-)$ are given in \eqref{S3coords} (with $\phi_- = \phi_1 - \phi_2$).  The brane web metric, \eqref{defbwmet}, can be put in the form
\begin{align}\label{defbwst}
& ds_{6}^2 = \ell_{s}^2 \left( \frac{\alpha^2}{4} d\vartheta^2 + \alpha^2 \sin^2{\frac{\vartheta}{2}} d\phi_{w}^2 + \alpha^{-2} \sec^2{\frac{\vartheta}{2}} d\phi_{7}^2 \right) + \OO(\tau^2)~, \qquad \textrm{with} \cr
& \alpha^2 = \left( \frac{|\vareps|^2}{\ell_{s} R_7} \right)^{2/3}~.
\end{align}
This result is precisely what one would expect from T-duality.  It can be obtained from the round metric on the three-sphere at the core of the deformed conifold by first presenting $S^3$ as a circle fibration over the disk, where the fibre direction is identified with the T-duality direction.  Then we simply invert the size of the circle fibre, sending the radius of the circle to $\ell_{s}^2$ over the radius.  In particular, the $\phi_7$ circle in \eqref{defbwst} blows up as we approach the brane locus at the boundary of the disk.

Finally, let us discuss the Neveu--Schwarz two-form, $B_2$.  The $B$-field is defined through \eqref{B2} and is patch dependent.    The patches are the same ones used in defining $\KK^{\rm reg}$: $\UU_+$ where $x \neq 0$, and $\UU_-$ where $u \neq 0$.  These conditions should be expressed in terms of our radial-angular coordinate system.  Using \eqref{defcov} we find
\begin{equation}\label{Bdefpatches}
\UU_{\pm}~: \qquad (1+ c_1 c_2) \cosh{\tau} \pm (c_1+c_2) \sinh{\tau} - s_1 s_2 \cos{\psi} \neq 0~.
\end{equation}
The set of points not covered by $\UU_+ \cup \UU_-$ is the brane locus, where of course the $B$-field can not be defined.  On $\UU_{\pm}$ we find
\begin{align}\label{B2def}
B_{2}^{\pm} =&~ -\frac{\ell_{s}^2}{2 F} \displaystyle\biggl\{ 6 K(\tau)^3 s_{\psi} (s_2 d\theta_1 + s_1 d\theta_2) - 4(c_1 + c_2) d\psi  + \cr
& \qquad \qquad \qquad   - \left( 2- 3 ch_{\tau} K(\tau)^3 \right) (c_{1}^2 - c_{2}^2) d\phi_{-}^2 + d \Lambda_{\pm} \displaystyle\biggr\} \wedge d\phi_7~,
\end{align}
where the pure gauge piece involves
\begin{equation}\label{Lambdadefpm}
\Lambda_{\pm} = \pm {\rm Arg} \left[ \cos{\frac{\theta_1}{2}} \cos{\frac{\theta_2}{2}} e^{\pm \half (\tau + i \psi)} - \sin{\frac{\theta_1}{2}} \sin{\frac{\theta_2}{2}} e^{\mp \half(\tau + i \psi)} \right]~.
\end{equation}
With the aid of \eqref{defcov} we may express these as $d \Lambda_+ = d (\phi_x - \phi_+)$, and $d \Lambda_- = - d (\phi_u + \phi_+)$, implying  $\Lambda_{+}(\Lambda_-)$ is well-defined precisely on the patch $\UU_+(\UU_-)$.  The difference between $B_{2}^{\pm}$ on the overlap $\UU_+ \cap \UU_-$ is pure gauge, as required for consistency of the background.

The gauge transformation relating $B_{2}^{\pm}$ on the overlap encodes the nonzero H-charge of the system and can be used to compute it.  We have that $d \Lambda_+ - d\Lambda_- = d(\phi_x + \phi_u)$, and since $x u = v w + \half \vareps^2$,
\begin{equation}\label{Bdefdiff}
B_{2}^+ - B_{2}^- = -\ell_{s}^2 \ d  {\rm Arg}(\lambda - \lambda_0) \wedge d\phi_7~.
\end{equation}
Now consider a three-cycle enclosing the brane web of the form $\Sigma_3 = S_{x^7}^1 \times \Sigma_2$, with $\Sigma_2$ a two-sphere in $(y,\lambda)$ space centred on $(0,\lambda_0)$.  We can choose the two-sphere such that on the upper hemisphere $B_{2}^+$ is well-defined and on the lower one $B_{2}^-$ is.  Let $\d H^+$ denote the boundary of the upper hemisphere.  We then have
\begin{equation}\label{defHcharge}
Q_0 = -  \int_{\Sigma_3} H_3 = - \int_{\d H^+ \times S_{x^7}^1} (B_{2}^+ - B_{2}^-) = (2\pi \ell_s)^2~.
\end{equation}
This is the the NS-charge of a single NS5-brane.

\vskip2cm
\noindent {\bf Acknowledgements:} It is a pleasure to thank our respective instutions for hospitality while this work was being completed. JM is supported by an EPSRC Fellowship EP/G051054/1. AR acknowledges support from DOE grant DE-FG02-96ER50959.
\vskip2cm

%%%%%%%%%%%%%%%%%%%%%%%%%%%%%%%%%%%%%%%%%%%%%%
%%%%%%%%%%%%%%%%%%%%%%%%%%%%%%%%%%%%%%%%%%%%%%
\appendix
%%%%%%%%%%%%%%%%%%%%%%%%%%%%%%%
%%%%%%%%%%%%%%%%%%%%%%%%%%%%%%%
\section{Proof of holomorphic profiles}
%%%%%%%%%%%%%%%%%%%%%%%%%%%%%%%
%%%%%%%%%%%%%%%%%%%%%%%%%%%%%%%

In this appendix we prove that $1/4$-BPS brane webs follow holomorphic profiles by studying the equation of motion for the flux in a near-brane limit, and demanding consistency with the parameterisation \eqref{NSfive} determined from supersymmetry.  For lower-dimensional brane webs we show that the equation of motion additionally implies that the induced metric on the brane worldvolume is flat, while for NS5-brane webs there are no further constraints beyond holomorphicity of the profile.  It is sufficient to consider the equation of motion for the flux since supersymmetry implies that the remaining equations of motion will be satisfied if one of them is.

%%%%%%%%%%%%%%%%%%%%%%%%%%%%%%%
\subsection{Flux equations of motion and sourced Monge--Ampere}
%%%%%%%%%%%%%%%%%%%%%%%%%%%%%%%

NS5-branes are fundamental sources for the magnetic dual of the NS-NS two-form, so we write the type II supergravity action in terms of magnetic variables, $\HH_7 = d \BB_6$, related to the electric ones by Hodge duality, $\HH_7 = e^{-2\varphi} \star H_3$ \cite{Duff:1994an}.  The relevant part of the bulk plus brane action is
\begin{equation}\label{SofB}
S[\BB_6] = - \frac{1}{4 \kappa_{10}^2} \int e^{2\varphi} \HH_7 \wedge \star \HH_7 - n \ \mu_{\rm NS5} \int_{\Sigma_6} P[\BB_6]~.
\end{equation}
There is also a bulk Chern-Simons term in general, but it makes no contribution for the class of brane configurations we consider.  $P[\BB_6]$ denotes the pullback of the six-form potential to the worldvolume of the brane, $\Sigma_6$.  We will consider a single brane web of charge $n$ located at ${\bf y} = 0$; the generalisation to multiple webs at different positions in ${\bf y}$ is trivial.  The brane worldvolume wraps $\mathbb{R}^{1,3}$ and is extended along a two-dimensional surface in the space spanned by $(z^a,\zbar^{\abar})$.  We introduce real local tangential coordinates $\eta_i$ and orthogonal coordinates $\lambda_i$, $i=1,2$, such that the brane embedding is described by $\lambda_i = 0$, while $\eta_i$ parameterise the worldvolume.

The $\HH_7$ consistent\footnote{See Appendix A of \cite{McOrist:2011in} for details.  Note that $H_7$ used there is $H_7 = e^{2\varphi} \HH_7$.} with \eqref{NSfive} is $\HH_7 = -i dt \wedge d^3{\bf x} \wedge d \left( \KK_{a\bbar} dz^a d\zbar^{\bbar} \right)$, and therefore we may assume the only nonzero legs of $\BB_6$ are of the form $(\BB_6)_{\mu\nu\rho\sigma a\bbar}$, up to antisymmetric permutation.  This allows us to express the pullback in the form
\begin{equation}\label{PBtoB}
\int_{\Sigma_6} P[\BB_6] = \int \BB_6 \wedge \delta^{(2)}(\lambda_1,\lambda_2) \delta^{(2)}({\bf y} ) d\lambda_1 d\lambda_2 d^2 {\bf y}~,
\end{equation}
where the integral on the right is over ten-dimensional spacetime.  We thus have the equation of motion
\begin{equation}\label{fluxeom}
d H_3 = -Q  \delta^{(2)}(\lambda_1,\lambda_2) \delta^{(2)}({\bf y} ) d\lambda_1 d\lambda_2 d^2 {\bf y}~,
\end{equation}
where $Q = 2 n \kappa_{10}^2 \mu_{\rm NS5} = (2\pi \ell_{s})^2n$.  On the other hand, from \eqref{NSfive} and \eqref{gAK}, we find that
\begin{equation}\label{dH3}
dH_3 = i \d_a \delbar_{\bbar} \left( \Delta_{\bf y} \KK + 8 \det{(\d \delbar \KK)} \right) dz^a d\zbar^{\bbar} d^2 {\bf y}~.
\end{equation}
We write $d\lambda_i = \d_a \lambda_i dz^a + \delbar_{\abar} \lambda_i d\zbar^{\abar}$ and plug into \eqref{fluxeom}.  Consistency of the brane embedding with supersymmetry, viz. \eqref{dH3}, implies
\begin{equation}\label{c1}
\varepsilon^{ab} \d_a \lambda_1 \d_b \lambda_2 \displaystyle\bigg|_{\Sigma_6} = 0~,
\end{equation}
and its conjugate.  Equating \eqref{fluxeom} with \eqref{dH3} for the remaining terms in the expansion of $d\lambda_1 d\lambda_2$ implies
\begin{equation}\label{sourcedMA}
 \d_a \delbar_{\bbar} \left( \Delta_{\bf y} \KK + 8 \det{(\d \delbar \KK)} \right) = i Q (\d_a \lambda_1 \delbar_{\bbar} \lambda_2 - \d_a \lambda_2 \delbar_{\bbar} \lambda_1 ) \delta^{(2)}(\lambda_1,\lambda_2) \delta^{(2)}({\bf y})~,
\end{equation}
with $\Delta_{\bf y}$ the (flat-space) Laplacian on $\mathbb{R}^2$.

Our goal in the remainder of this section is to understand the content of \eqref{sourcedMA}---in particular, the additional constraints beyond \eqref{c1} that must be imposed on $\lambda_i(z^a,\zbar^{\abar})$.  We will consider a slight generalisation of \eqref{sourcedMA}, where we allow ${\bf y}$ to span a $d$-dimensional transverse space,
\begin{equation}\label{sourcedMAd}
 \d_a \delbar_{\bbar} \left( \Delta_{\bf y} \KK + 8 \det{(\d \delbar \KK)} \right) = -i Q (\d_a \lambda_1 \delbar_{\bbar} \lambda_2 - \d_a \lambda_2 \delbar_{\bbar} \lambda_1 ) \delta^{(2)}(\lambda_1,\lambda_2) \delta^{(d)}({\bf y})~.
\end{equation}
The $d=2$ case corresponds to (localised) NS5-brane webs while $d=5$ corresponds to membrane webs, the case originally considered in \cite{Lunin:2008tf}.  We will eventually be interested in NS5-brane webs smeared on a transverse circle, corresponding to $d=1$.  If we can demonstrate that localised NS5-brane webs follow holomorphic profiles, however, then by construction so will smeared ones.

%%%%%%%%%%%%%%%%%%%%%%%%%%%%%
\subsection{The near-brane limit}
%%%%%%%%%%%%%%%%%%%%%%%%%%%%%

Our strategy will be to analyse the singularity structure of \eqref{sourcedMAd} in a near-brane limit where we approach the source locus.  More precisely, we set
\begin{equation}\label{nblimit}
\lambda_{i} = \epsilon\hat{\lambda}_i ~, \qquad {\bf y} = \epsilon\hat{{\bf y}}~,
\end{equation}
where $\eta_i,\hat{\lambda}_i,\hat{{\bf y}}$ are $\OO(1)$, and study the behaviour of \eqref{sourcedMAd} around $\epsilon = 0$.  First let us extract some useful information from the condition \eqref{c1} which, in this language, states
\begin{equation}\label{vzJac}
\d_1 \lambda_1 \d_2 \lambda_2 - \d_1 \lambda_2 \d_2 \lambda_1 = \OO(\epsilon)~.
\end{equation}
It follows from this that the Jacobian for the change of variables $(z^a,\zbar^{\abar}) \mapsto (\lambda_i, \eta_i)$ takes a factored form as we approach the locus:
\begin{equation}\label{Jacfac}
\displaystyle\bigg| \frac{ \d(\lambda_i,\eta_i) }{\d(z^a,\zbar^\abar) } \displaystyle\bigg| =   \varepsilon^{ab} \varepsilon^{\abar\bbar}  (\d_a \lambda_1 \delbar_{\abar} \lambda_2 - \d_a \lambda_2 \delbar_{\abar} \lambda_1 ) ( \d_b \eta_1 \delbar_{\bbar} \eta_2  - \d_b \eta_2 \delbar_{\bbar} \eta_1) + \OO(\epsilon)~.
\end{equation}
Here and below we restrict ourselves to smooth brane embeddings, so that $\d_a \lambda_i, \d_a \eta_i$, and higher order derivatives never diverge with $\epsilon$.

Now let us determine the degree of divergence of $\KK$.  First we note that the Jacobian-like factor on the right side of \eqref{sourcedMAd} must be $\OO(1)$ for at least one $(a,\bbar)$ pair.  If it were $\OO(\epsilon)$ or smaller for all values of $(a,\bbar)$, the Jacobian for the change of variables, \eqref{Jacfac}, would be $\OO(\epsilon)$ or smaller.  The Jacobian would vanish on the brane locus, but this is a contradiction since we assume that $(\eta_i,\lambda_i)$ is a good coordinate system in a neighbourhood of the locus.  Thus the right side of \eqref{sourcedMAd} diverges as $\OO(\epsilon^{-(d+2)})$ for at least one $(a,\bbar)$ pair.

On the left side of \eqref{sourcedMAd} we change variables from $(z^a,\zbar^{\bbar})$ to $(\lambda_i,\eta_i)$; in particular, $\d_a \delbar_{\bbar} = (\d_a \lambda_i \delbar_{\bbar} \lambda_j) \d_{\lambda_i} \d_{\lambda_j} + \cdots$, summing over $i,j$.  Derivatives with respect to $\lambda_i$ and ${\bf y}$ bring one power of $\epsilon^{-1}$ each.  Now for the same $(a,\bbar)$ pair of the previous paragraph, we have that $\d_a \lambda_i, \delbar_{\bbar} \lambda_i = \OO(1)$, and thus we are guaranteed that $\d_a \delbar_{\bbar} = \OO(\epsilon^{-2})$.  Matching leading divergences on the left and right then implies $\left( \Delta_{\bf y} \KK + 8 \det{(\d \delbar \KK)} \right) = \OO(\epsilon^{-d} )$.  On physical grounds, the potential $\KK$ should not itself contain a Dirac delta function in ${\bf y}$.  Therefore it must be the $\Delta_{\bf y} \KK$ term that is responsible for generating the right-hand side of \eqref{sourcedMAd}.  We conclude that
\begin{align}\label{Korder}
\KK = \OO(\epsilon^{2-d})~, \qquad d \geq 2~.
\end{align}
For now we only keep track of the leading divergence of $\KK$; there may also be pieces that have a subleading divergence, or pieces that are regular on the brane locus.  We also restrict to $d \geq 2$ until further notice.  When $d=2$, \eqref{Korder} should be understood as a log divergence in $\epsilon$.

The fact that the leading divergence of $\KK$ is $\OO(\epsilon^{2-d})$ has interesting implications for the determinant term in \eqref{sourcedMAd}.  Since this term may contain up to four $\lambda_i$-derivatives, we would naively conclude that $\det{(\d \delbar \KK)} = \OO( \epsilon^{-2d})$.  However, the right side of \eqref{sourcedMAd} implies that it can be no more divergent than $\epsilon^{-d}$.  Therefore it must be that $\det{(\d \delbar \KK)}$ vanishes at each order $\epsilon^{-n}$ for $d < n \leq 2d$.  We next turn to a systematic investigation of these constraints.

%%%%%%%%%%%%%%%%%%%%%%%%%%%
\subsection{$\epsilon$ expansion of the determinant}
%%%%%%%%%%%%%%%%%%%%%%%%%%%

The $2\times 2$ matrix we are taking the determinant of has matrix elements
\begin{align}\label{Kaborg}
\d_a \delbar_{\bbar} \KK =&~ \d_a \lambda_i \delbar_{\bbar} \lambda_j \KK_{\lambda_i \lambda_j} + \cr
& + (\d_a \delbar_{\bbar} \lambda_i) \KK_{\lambda_i} + ( \d_a \lambda_i \delbar_{\bbar} \eta_j + \d_a \eta_j \delbar_{\bbar} \lambda_i ) \KK_{\lambda_i \eta_j} + \cr
& + (\d_a \delbar_{\bbar} \eta_i) \KK_{\eta_i} + \d_a \eta_i \delbar_{\bbar} \eta_j \KK_{\eta_i \eta_j}~,
\end{align}
where we use the shorthand $f_{\lambda_i} \equiv \d_{\lambda_i} f$.  Each line represents a decreasing degree of divergence, with the top line going as $\epsilon^{-d}$.  Strictly speaking, each term displayed is potentially this order, but may be subleading if a particular $\d_a \lambda_i$ vanishes on the brane locus, or if we are evaluating on a subdivergent piece of $\KK$.

Using the expansion \eqref{Kaborg}, we have the following potential contribution to $\det{(\d \delbar \KK)}$ at $\OO(\epsilon^{-2d})$:
\begin{align}\label{leadingdet}
\det{(\d \delbar \KK)} =&~ \left| \begin{array}{c c} \d_{1} \lambda_i \delbar_{\bar{1}} \lambda_j \KK_{\lambda_i \lambda_j} & \d_{1} \lambda_i \delbar_{\bar{2}} \lambda_j \KK_{\lambda_i \lambda_j} \\ \d_{2} \lambda_k \delbar_{\bar{1}} \lambda_l \KK_{\lambda_k \lambda_l} & \d_{2} \lambda_k \delbar_{\bar{2}} \lambda_l \KK_{\lambda_k \lambda_l} \end{array} \right| + \OO(\epsilon^{1-2d}) \cr
=&~ \d_{1} \lambda_i \delbar_{\bar{1}} \lambda_j  \d_{2} \lambda_k \delbar_{\bar{2}} \lambda_l \left[ \KK_{\lambda_i \lambda_j} \KK_{\lambda_k \lambda_l} - \KK_{\lambda_i \lambda_l} \KK_{\lambda_j \lambda_k} \right] + \OO(\epsilon^{1-2d})~.
\end{align}
Observe that in \eqref{leadingdet} we must sum over all combinations $(ijkl)$ where each index can take the value one or two.  However, most of these terms vanish trivially due to vanishing of the square bracketed term.  The only cases that don't vanish are $(ijkl) = (1122), (2211),(1221)$, and $(2112)$.  These terms combine to give
\begin{align}\label{leadingdet2}
\det{(\d \delbar \KK)} =&~ \left[ \KK_{\lambda_1 \lambda_1} \KK_{\lambda_2 \lambda_2} - \KK_{\lambda_1 \lambda_2}^2 \right] \left| \d_1 \lambda_1 \d_2 \lambda_2 - \d_2 \lambda_1 \d_1 \lambda_2 \right|^2 + \OO(\epsilon^{1-2d})~.
\end{align}
The second term in this factor vanishes to $\OO(\epsilon^2)$ thanks to \eqref{vzJac}.  Thus, compatibility of the flux equation of motion with supersymmetry (which gave us \eqref{vzJac}) already guarantees that the $\OO(\epsilon^{-2d})$ contribution to the determinant vanishes.

Next we consider the potential contributions at $\OO(\epsilon^{1-2d})$.  Since we are considering $d \geq 2$, these contributions must vanish according to \eqref{sourcedMAd}.  They come from cross-terms in the determinant when terms from the first line of \eqref{Kaborg} hit terms from the second line.  It will be convenient to introduce some notation; let
\begin{align}\label{effs}
& f_{a\bbar} = f_{a\bbar}^{(1)} + f_{a\bbar}^{(2)}~, \qquad \textrm{with} \cr
& f_{a\bbar}^{(1)} = (\d_a \delbar_{\bbar} \lambda_i) \KK_{\lambda_i}~, \qquad f_{a\bbar}^{(2)} = ( \d_a \lambda_i \delbar_{\bbar} \eta_j + \d_a \eta_j \delbar_{\bbar} \lambda_i ) \KK_{\lambda_i \eta_j}~.
\end{align}
Then
\begin{align}\label{subleadingdet}
\det{(\d \delbar \KK)} =&~ \left| \begin{array}{c c} \d_{1} \lambda_i \delbar_{\bar{1}} \lambda_j \KK_{\lambda_i \lambda_j} + f_{1\bar{1}} & \d_{1} \lambda_i \delbar_{\bar{2}} \lambda_j \KK_{\lambda_i \lambda_j} + f_{1\bar{2}} \\ \d_{2} \lambda_k \delbar_{\bar{1}} \lambda_l \KK_{\lambda_k \lambda_l} + f_{2\bar{1}} & \d_{2} \lambda_k \delbar_{\bar{2}} \lambda_l \KK_{\lambda_k \lambda_l} + f_{2\bar{2}} \end{array} \right| + \OO(\epsilon^{2-2d}) \cr
=&~ f_{2\bar{2}} \d_{1} \lambda_i \delbar_{\bar{1}} \lambda_j \KK_{\lambda_i \lambda_j} + f_{1\bar{1}} \d_{2} \lambda_i \delbar_{\bar{2}} \lambda_j \KK_{v_i v_j}  + \cr
& \qquad \quad - f_{2\bar{1}} \d_{1} \lambda_i \delbar_{\bar{2}} \lambda_j \KK_{\lambda_i \lambda_j} - f_{1\bar{2}} \d_{2} \lambda_i \delbar_{\bar{1}} \lambda_j \KK_{\lambda_i \lambda_j} + \OO(\epsilon^{2-2d})~.
\end{align}

We know that for at least one $(a,\bbar)$ pair, $\d_a \lambda_i, \delbar_{\bbar} \lambda_i = \OO(1)$.  Suppose this pair is $(1,\bar{1})$.  Then, expanding out the sums over $i,j$ in \eqref{subleadingdet}, we find that the terms may be collected as follows:
\begin{align}\label{FandL}
& \det{(\d \delbar \KK)} = \left[ F_{(1,\bar{1})}^{(1)} + F_{(1,\bar{1})}^{(2)} \right] L_{2}^{(1,\bar{1})}[\KK] + \OO(\epsilon^{2-2d})~, \qquad \textrm{with} \cr
& F_{(1,\bar{1})}^{(i)} =  f_{2\bar{2}}^{(i)} + f_{1\bar{1}}^{(i)} \frac{ |\d_2 \lambda_1|^2}{| \d_1 \lambda_1|^2} - f_{2\bar{1}}^{(i)} \frac{ \delbar_{\bar{2}} \lambda_1}{ \delbar_{\bar{1}} \lambda_1} - f_{1\bar{2}}^{(i)} \frac{ \d_2 \lambda_1}{\d_1 \lambda_1} ~, \qquad i = 1,2, \cr
& L_{2}^{(1,\bar{1})}[\KK] =  |\d_1 \lambda_1|^2 \KK_{\lambda_1 \lambda_1} +  |\d_1 \lambda_2|^2 \KK_{\lambda_2 \lambda_2} + (\d_1 \lambda_1 \delbar_{\bar{1}} \lambda_2 + \d_1 \lambda_2 \delbar_{\bar{1}} \lambda_1) \KK_{\lambda_1 \lambda_2}~.
\end{align}
We arrived at this expression by collecting the coefficients in front of each $\KK_{\lambda_i \lambda_j}$ in \eqref{subleadingdet}, and then dividing through by appropriate factors of $\d_1 \lambda_i, \delbar_{\bar{1}} \lambda_i$ so that, using \eqref{vzJac}, all of the resulting coefficients are the same to leading order and given by $F_{(1,\bar{1})} = F_{(1,\bar{1})}^{(1)} + F_{(1,\bar{1})}^{(2)}$.  If the $(a,\bbar)$ pair is one of the other three possibilities, we can derive completely analogous expressions to \eqref{FandL}, by dividing through by $\d_a \lambda_i$ and $\delbar_{\bbar} \lambda_i$ to create common coefficients.  The remainder of the analysis is technically identical in each case, so we will restrict to $(a,\bbar) = (1,\bar{1})$ to avoid overcomplicating the discussion.

Let us study the $F_{(1,\bar{1})}^{(i)}$ using \eqref{effs}.  After some rearranging we find that
\begin{align}\label{F2}
F_{(1,\bar{1})}^{(2)} =&~ \left[ \left( \delbar_{\bar{2}} \eta_j - \frac{\delbar_{\bar{2}} \lambda_1}{ \delbar_{\bar{1}} \lambda_1} \delbar_{\bar{1}} \eta_j \right) \left( \d_2 \lambda_i - \d_1 \lambda_i \frac{\d_2 \lambda_1}{\d_1 \lambda_1} \right) + c.c. \right] \KK_{v_i w_j} \cr
=&~ \OO(\epsilon)~.
\end{align}
In the second step we noted that the $\lambda$ term vanishes trivially for $i=1$, while for $i=2$ we can use \eqref{vzJac}.  Similar manipulations lead to
\begin{equation}\label{F1}
F_{(1,\bar{1})}^{(1)} = \left[ \left( \d_2 - \frac{ \d_2 \lambda_1}{\d_1 \lambda_1} \d_1 \right) \left( \frac{ \delbar_{\bar{2}} \lambda_1}{ \delbar_{\bar{1}} \lambda_1} \right) \right] (\delbar_{\bar{1}} \lambda_i) \KK_{\lambda_i} + \OO(\epsilon)~.
\end{equation}
It follows that
\begin{equation}\label{subleadingdet2}
\det{(\d \delbar \KK)} =  \left[ \left( \d_2 - \frac{ \d_2 \lambda_1}{\d_1 \lambda_1} \d_1 \right) \left( \frac{ \delbar_{\bar{2}} \lambda_1}{ \delbar_{\bar{1}} \lambda_1} \right) \right] \cdot (\delbar_{\bar{1}} \lambda_i) \KK_{\lambda_i} \cdot L_{2}^{(1,\bar{1})}[\KK]  + \OO(\epsilon^{2-2d})~.
\end{equation}

To summarise where we are, \eqref{subleadingdet2} is the most divergent part of $\det{(\d \delbar \KK)}$.  All terms which are potentially as divergent or more divergent have vanished using the condition \eqref{vzJac}.  \eqref{subleadingdet2} is the first potentially new condition that we have to work with.  Superficially, it diverges as $\OO(\epsilon^{1-2d})$, where the counting goes as follows.  First, $L_{2}^{(1,\bar{1})}[\KK]$ goes as $\OO(\epsilon^{-d})$ since it involves two $\lambda_i$-derivatives on $\KK$, and the prefactors $\d_1 \lambda_i, \delbar_{\bar{1}} \lambda_i$ are guaranteed to be $\OO(1)$.  The only way this counting could fail is if the terms in $L_{2}^{(1,\bar{1})}[\KK]$ cancel among themselves at leading order.  Similarly, the factor $(\delbar_{\bar{1}} \lambda_i) \KK_{\lambda_i}$ is superficially $\OO(\epsilon^{1-d})$, since it involves one $\lambda_i$-derivative acting on $\KK$, and the prefactor is guaranteed to be $\OO(1)$.  Finally, for generic embedding functions $\lambda_i(z^a,\zbar^{\bbar})$, one expects the square-bracketed term coming from $F_{(1,\bar{1})}^{(1)}$ to be $\OO(1)$.

Since $\epsilon^{1-2d}$ is more divergent than $\epsilon^{-d}$ for $d \geq 2$, it must in fact be that \eqref{subleadingdet2} is less divergent than this naive counting suggests.  This requires that one of the three factors be subleading to the naive expectation for it.  We can immediately rule this out for $(\delbar_{1} \lambda_i) \KK_{\lambda_i}$.  Suppose it is the case that $(\delbar_{\bar{1}} \lambda_i) \KK_{\lambda_i} = \OO(\epsilon^{1-d+n})$, where $n>0$ is the degree by which the expression is subdivergent to naive expectations.  Then, since $\KK_{\lambda_i}$ is real,
\begin{align}\label{K2lambdai}
\delbar_{\bar{1}} \lambda_1 \KK_{\lambda_1} + \delbar_{\bar{1}} \lambda_2 \KK_{\lambda_2} = \OO(\epsilon^{2-d+n}) \quad \Rightarrow  \quad &  \frac{ \delbar_{\bar{1}} \lambda_1}{\delbar_{\bar{1}} \lambda_2} = - \frac{\KK_{\lambda_2}}{ \KK_{\lambda_1}} + \OO(\epsilon^n) = \frac{\d_1 \lambda_1}{\d_1 \lambda_2} \cr
 \Rightarrow \quad & \delbar_{\bar{1}} \lambda_1 \d_1 \lambda_2 - \delbar_{\bar{1}} \lambda_2 \d_1 \lambda_1 = \OO(\epsilon^{n})~.
\end{align}
However, if this is true, then using \eqref{vzJac} and its conjugate, we can derive
\begin{align}\label{nottrue}
 \Rightarrow \quad & \d_a \lambda_1 \delbar_{\bbar} \lambda_2 - \d_a \lambda_2 \delbar_{\bbar} \lambda_1 = \OO(\epsilon^n)~,
\end{align}
for any $(a,\bbar)$ pair.  As discussed around \eqref{Jacfac}, this would imply that $(\eta_i,\lambda_i)$ is not a good coordinate system around the brane locus--a contradiction.

We can also argue against $L_{2}^{(1,\bar{1})}[\KK_2]$ being subdivergent to the naive expectation.  As we know from our discussion above \eqref{Korder}, it is the $\Delta_{\bf y} \KK$ term on the left side of \eqref{sourcedMAd} that is responsible for generating the source term.  However, consider the $(a,\bbar) = (1,\bar{1})$ component of the left side:
\begin{align}\label{ddbarDelK}
\d_1 \delbar_{\bar{1}} \Delta_{\bf y} \KK =&~ \Delta_{\bf y} (\d_1 \delbar_{\bar{1}} \KK) = \Delta_{\bf y} \left[ \d_1 \lambda_i \delbar_{\bar{1}} \lambda_j \KK_{\lambda_i \lambda_j} + \cdots \right] \cr
=&~ \Delta_{\bf y} \left( L_2^{(1,\bar{1})}[\KK] + \cdots \right)~.
\end{align}
Thus we must have $L_{2}^{(1,\bar{1})}[\KK] = \OO(\epsilon^{-d})$ in order to generate a source term of the correct order.

The only remaining possibility for the $\OO(\epsilon^{1-2d})$ divergence of \eqref{subleadingdet2} to vanish is that
\begin{equation}\label{c2}
 \left( \d_2 - \frac{ \d_2 \lambda_1}{\d_1 \lambda_1} \d_1 \right) \left( \frac{ \delbar_{\bar{2}} \lambda_1}{ \delbar_{\bar{1}} \lambda_1} \right) = \OO(\epsilon)~.
\end{equation}
Next we demonstrate that this condition, in conjunction with our other constraints on $\lambda_i(z^a,\zbar^{\abar})$, implies holomorphic profiles.

%%%%%%%%%%%%%%%%%%%%%%%%
\subsection{Holomorphic profiles}
%%%%%%%%%%%%%%%%%%%%%%%%

Clearly one class of solutions to \eqref{c2} is that $\d_2 \lambda_1/\d_1 \lambda_1 = c(z^a)$ to leading order, where $c(z^a)$ is an arbitrary holomorphic function.  Using \eqref{vzJac}, this implies the same relation for $\d_2 \lambda_2/\d_1 \lambda_2$, so within this class we have
\begin{equation}\label{specialsol}
\d_2 \lambda_i - c(z^a)  \d_1 \lambda_i = \OO(\epsilon)~, \qquad i = 1,2.
\end{equation}
We solve this equation by the method of characteristics.  Consider curves in the $z^1$-$z^2$ plane satisfying
\begin{equation}\label{ccurve}
\frac{dz^1}{dz^2} = -c(z^1,z^2) + \OO(\epsilon)~.
\end{equation}
Integrating this equation we find
\begin{equation}\label{ccurve2}
z^1 = z^1(z^2;\alpha) + \OO(\epsilon)~,
\end{equation}
where $\alpha$ is an integration constant whose value parametrises a one-parameter family of curves.  On any one of these curves we have that $\lambda_i$ is constant to the order we are working:
\begin{equation}\label{liconstant}
\frac{d}{dz^2} \lambda_i \left( z^1(z^2;\alpha), z^2 ; \zbar^{\abar} \right)  = \d_1 \lambda_i \frac{d z^1}{dz^2} + \d_2 \lambda_i = \OO(\epsilon)~.
\end{equation}
It follows that $\lambda_i$ should only depend on $z^a$ through a particular function whose level sets give the characteristic curves.  This function is found by solving \eqref{ccurve2} for $\alpha$:
\begin{equation}\label{asolve}
z^1 = z^1(z^2;\alpha) + \OO(\epsilon) \quad \Rightarrow \qquad \alpha = \lambda(z^a) + \OO(\epsilon)~.
\end{equation}
We have labelled the leading order in $\epsilon$ part of the resulting function $\lambda$, which is a holomorphic function of the $z^a$.  Thus we conclude that
\begin{equation}\label{lispecialsol}
\lambda_i = \lambda_i\left( \lambda(z^a) + \OO(\epsilon); \zbar^{\abar} \right)~.
\end{equation}
Now, we could have equally well analysed the conjugate of equation \eqref{specialsol}.  Doing so, we determine the $\zbar^{\abar}$ dependence of $\lambda_i$:
\begin{equation}\label{lispecialsol2}
\lambda_i = \lambda_i \left( \lambda(z^a) + \OO(\epsilon); \lambdabar(\zbar^{\abar}) + \OO(\epsilon) \right) = \lambda_i (\lambda(z^a),\lambdabar(\zbar^{\abar})) + \OO(\epsilon)~,
\end{equation}
where $\lambdabar(\zbar^{\abar})$ is the conjugate of $\lambda(z^a)$.  But this is exactly the result we are after.  We can view equation \eqref{lispecialsol2} as the statement that there exists a change of coordinates which, when restricted to the brane locus, takes the form $(\lambda_1,\lambda_2) \mapsto (\lambda,\lambdabar)$, where $\lambda(z^a)$ is a holomorphic function of $z^a$.  By choosing boundary conditions for the first order PDE \eqref{specialsol} appropriately, we may assume that $\lambda_i = 0$ corresponds to $\lambda = \lambdabar = 0$.  Hence the brane locus is described by the holomorphic equation $\lambda(z^a) = 0$.

In order to complete the proof, we must demonstrate that any other solution to \eqref{c2} outside of the class \eqref{specialsol} is inconsistent with our other results concerning $\lambda_i(z^a,\zbar^{\abar})$.  Thus we consider
\begin{equation}\label{cgen}
\d_2 \lambda_i - c(z^a,\zbar^{\abar}) \d_1 \lambda_i = \OO(\epsilon)~, \qquad i = 1,2,
\end{equation}
where $c$ is a non-trivial function of both $z^a, \zbar^{\abar}$.  The function $c$ and its complex conjugate $\cbar(z^a,\zbar^{\abar})$ are required to satisfy the coupled equation
\begin{equation}\label{ccoupled}
(\d_2 - c\ \d_1) \cbar = \OO(\epsilon)~.
\end{equation}
Suppose we have such a solution $c$.  Then we may solve \eqref{cgen} using the method of characteristics as before.  This time, however, when we solve for $\alpha$ it will be a function of both the $z^a$ and their conjugates, and thus
\begin{equation}\label{ligenc}
\lambda_i = \lambda_i \left( \alpha(z^a; \zbar^{\abar}) ; \zbar^{\bbar} \right) + \OO(\epsilon)~.
\end{equation}
Meanwhile, solving the conjugate of \eqref{cgen} leads to
\begin{equation}\label{ligencc}
\lambda_i = \lambda_i \left(z^b; \alphabar(z^a; \zbar^{\abar}) \right) + \OO(\epsilon)~,
\end{equation}
where $\alphabar$ is the conjugate of $\alpha$.  The only way \eqref{ligenc} and \eqref{ligencc} can be consistent with each other is if $\alpha = \alphabar$ and
\begin{equation}\label{liareal}
\lambda_i = \lambda_i \left( \alpha(z^a;\zbar^{\abar}) \right) + \OO(\epsilon)~.
\end{equation}
If, however, $\lambda_1,\lambda_2$ depend on $(z^a,\zbar^\abar)$ through the same function, $\alpha$, we are in trouble since then
\begin{equation}\label{Jaccheck}
\d_a \lambda_1 \delbar_{\bbar} \lambda_2 - \d_a \lambda_2 \delbar_{\bbar} \lambda_1 = \lambda_1' \lambda_2' (\d_a \alpha \delbar_{\bbar}\alpha - \d_a \alpha \delbar_{\bbar} \alpha) + \OO(\epsilon) = \OO(\epsilon)~,
\end{equation}
for all pairs $(a,\bbar)$.  This implies that the Jacobian \eqref{Jacfac} will be $\OO(\epsilon)$, in contradiction to $(\lambda_i,\eta_i)$ being a good coordinate system in a neighbourhood of the locus.

%%%%%%%%%%%%%%%%%%%%%%%%%%%%%%%%%
\subsection{Flat induced metric on the worldvolume for $d > 2$}
%%%%%%%%%%%%%%%%%%%%%%%%%%%%%%%%%

Having established the existence of coordinates $(\lambda,\lambdabar)$, we can immediately put them to use.  First, we can simplify the right-hand side of \eqref{sourcedMAd} by changing coordinates from $(\lambda_1,\lambda_2)$ to $(\lambda,\lambdabar)$:
\begin{align}\label{sourcedMAd2}
& \d_a \delbar_{\bbar} \left( \Delta_{\bf y} \KK + 8 \det{(\d \delbar \KK)} \right) = -Q \d_a \lambda \delbar_{\bbar} \lambdabar \ \delta^{(d)}({\bf y}) \delta(\lambda,\lambdabar)~, \cr
\Rightarrow \quad & \Delta_{\bf y} \KK + 8 \det{(\d \delbar \KK)} = - \frac{Q}{2\pi} \delta^{(d)}({\bf y}) \log{|\lambda|^2}~.
\end{align}
(Here we used the freedom of shifting $\KK$ by the real part of an arbitrary holomorphic function to set integration constants to zero).  Second, we can construct a holomorphic coordinate $\eta(z^a)$ orthogonal to $\lambda$ by demanding that the Jacobian for the change of variables $(z^1,z^2) \mapsto (\eta, \lambda)$ has unit determinant, as we approach the locus.  Thus $(\eta,\etabar)$ will parameterise the brane worldvolume.  This allows us to simplify our near-brane analysis.  In particular, by changing coordinates $(z^a,\zbar^{\abar}) \mapsto (\eta,\lambda,\etabar,\lambdabar)$, we have
\begin{equation}\label{detsimp}
\det{(\d \delbar \KK)} = \left(  \KK_{\eta\etabar} \KK_{\lambda\lambdabar} - \KK_{\eta\lambdabar} \KK_{\lambda \etabar}  \right)  + \OO(\epsilon^{3-2 d})~.
\end{equation}
If $d = 2$, then the displayed term is order $\epsilon^{2-2d} = \epsilon^{-2}$.  This is the same order as the other terms in \eqref{sourcedMAd2}, and there are naively\footnote{It may be possible to refine this analysis by considering an expansion in degrees of logarithmic divergence.} no more constraints to be imposed.

If $d >2$ on the other hand, we must demand that the determinant \eqref{detsimp} vanish to a subleading order.  Let us define the order one quantities $\hlambda$, $\hat{\KK}$ such that $\lambda = \epsilon \hlambda$ and $\KK(\eta,\etabar,\lambda,\lambdabar,{\bf y}) = \epsilon^{2-d} \hat{\KK}(\eta,\etabar,\hlambda,\hlambdabar,\hat{{\bf y}})$.  (We choose boundary conditions for the first order PDE \eqref{specialsol} so that the $\lambda_i$ vanish linearly with $\lambda$.)  Then, since the determinant can be no more divergent than $\epsilon^{-d}$, we find the condition
\begin{equation}\label{detcon}
 \hat{\KK}_{\eta\etabar}  \hat{\KK}_{\hlambda \hlambdabar} - \hat{\KK}_{\eta \hlambdabar}  \hat{\KK}_{\hlambda \etabar} = \OO(\epsilon^{d-2})~.
\end{equation}
The quantity on the left that is naively order one must, in fact, vanish as $\epsilon^{d-2}$.  This equation, without the $\OO(\epsilon^{d-2})$ corrections, is precisely the sort of equation considered in \cite{Lunin:2008tf}.  There it was argued that it implies the existence of a holomorphic coordinate $\hgamma = \hgamma(\eta,\hlambda)$, such that $\hat{\KK}$ depends on $(\hgamma,\hgammabar,\hat{{\bf y}})$ only.  This type of nonlinear PDE has also been well studied in the math literature; in particular, the result of \cite{Lunin:2008tf} follows straightforwardly from Theorem 2.4 of \cite{BedfordKalka}.  These results imply the split
\begin{equation}\label{Khatsplit}
\hat{\KK} = \hat{\KK}^{\rm div}(\hgamma(\eta,\hlambda), \hgammabar(\etabar,\hlambdabar), \hat{{\bf y}}) + \hat{\KK}^{\rm fin}(\eta,\hlambda,\etabar,\hlambdabar, \hat{{\bf y}} )~,
\end{equation}
where $\hat{\KK}^{\rm div} = \OO(1)$ and $\hat{\KK}^{\rm fin} = \OO(\epsilon^{d-2})$.  After transcribing back to unhatted quantities, $\KK^{\rm div}$ diverges as $\epsilon^{2-d}$, while $\KK^{\rm fin}$ is finite on the locus.  It must be that $\Delta_{\bf y} \KK^{\rm div}$ generates the source term on the right-hand side of \eqref{sourcedMAd2}, and therefore we may identify the function $\gamma(\eta,\lambda) = \lambda$.  Hence \eqref{detcon} implies
\begin{align}\label{Ksplit}
&\KK = \KK^{\rm div}(\lambda,\lambdabar,{\bf y}) + \KK^{\rm fin}(\eta,\lambda,\etabar,\lambdabar,{\bf y})~, \qquad \textrm{with} \cr
& \KK^{\rm div} = \OO(\epsilon^{2-d})~, \qquad \KK^{\rm fin} = \KK^{\rm fin}(\eta,\etabar) + \OO(\epsilon)~, \qquad (d > 2)~.
\end{align}
With this split we finally have $\det{(\d \delbar \KK)} = \KK_{\eta\etabar}^{\rm fin} \ \KK_{\lambda \lambdabar}^{\rm div}$ plus order one corrections, with $ \KK_{\eta\etabar}^{\rm fin} \ \KK_{\lambda \lambdabar}^{\rm div} = \OO(\epsilon^{-d})$, so all constraint conditions on $\KK$ are satisfied.

In fact, this parameterisation of $\KK$ allows us to solve \eqref{sourcedMAd2} perturbatively in $\epsilon$.  At $\OO(\epsilon^{-d})$ we get a linear PDE for $\KK^{\rm div}$:
\begin{equation}\label{Poissond}
\left[ \Delta_{\bf y} + 8 \KK_{\eta\etabar}^{\rm fin}(\eta,\etabar) \ \d_\lambda \delbar_{\lambdabar} \right] \KK^{\rm div}(\lambda,\lambdabar,{\bf y}) = - \frac{Q}{2\pi} \delta^{(d)}({\bf y}) \log{|\lambda|^2}~, \qquad (d > 2)~.
\end{equation}
Note that only the leading piece of $\KK^{\rm fin}$ contributes at this order in $\epsilon$.  Since the leading piece is independent of $(\lambda,\lambdabar,{\bf y})$ we may treat this factor a constant when solving \eqref{Poissond}.  Since $\KK^{\rm div}$ does not depend on $(\eta,\etabar)$ however, consistency of \eqref{Poissond} implies that $\KK_{\eta\etabar}^{\rm fin}$ is a pure constant at leading order.  In other words, the induced metric on the brane worldvolume, $\KK_{\eta\etabar} |_{\epsilon = 0}$, is flat for $d > 2$.

The constancy of $\KK_{\eta\etabar}$ at leading order implies that the the leading behaviour of the warp factor, $e^{-3A} = \det{(\d \delbar \KK)}$, is independent of the tangential coordinate $\eta$.  This is the initial assumption on which the near-brane analysis in \cite{Lunin:2008tf} is based.  Here we have shown (for $d > 2$) that it is rather a result that can be derived from consistency of the sourced equation of motion.

When $d=2$ the situation is complicated due to the necessity of keeping track of different degrees of logarithmic divergence.  Note that when $d > 2$ one should also expect $\log{\epsilon}$ corrections at each order in $\epsilon$ since the divergence on the right-hand side of \eqref{sourcedMAd2} is  actually $\epsilon^{-d} \log{\epsilon}$, but the analysis did not require us to explicitly keep track of $\OO(\epsilon^n)$ and $\OO(\epsilon^n \log{\epsilon})$ terms separately.  When $d = 2$ on the other hand, one expects the leading divergence of $\KK$ to be order $(\log{\epsilon})^2$.  This is the unique case where the divergent part of $\KK$ and the regular part differ by powers of $\log{\epsilon}$ only.  We know that derivatives with respect to $\lambda,{\bf y}$ go as $\epsilon^{-1}$ while derivatives with respect to $\eta$ go as $\epsilon^0$, but these derivatives may or may not cancel powers of $\log{\epsilon}$, depending on the detailed nature of the function $\KK$.  This ambiguity makes it difficult to perform the type of analysis done above for $d > 2$.

When $d=1$, corresponding to NS5-brane webs smeared on a transverse circle, our methods can again be used to analyse \eqref{sourcedMAd2}.  In this case the part of $\KK$ that should generate the source term is $\OO(\epsilon)$ and subleading to the regular piece of $\KK$ that is finite on the locus.  We write
\begin{equation}\label{Ksplitd1}
\KK = \KK_1(\eta,\etabar) + \KK_2(\eta,\lambda,\etabar,\lambdabar,y) + \OO(\epsilon^2)~, \qquad (d = 1)~,
\end{equation}
where $\KK_1= \OO(1)$ and $\KK_2 = \OO(\epsilon)$.  Here $\KK_2$ may contain both order $\epsilon$ and $\epsilon \log{\epsilon}$ terms.  It must contain a term whose second derivatives are singular, but it may also contain terms with regular derivatives.  Unlike the $d > 2$ case, there is no constraint on $\det{(\d \delbar \KK_2)}$, which goes as $\OO(\epsilon^0)$ and is subleading to the source term in \eqref{sourcedMAd2}.  Thus $\KK_2$ is free to depend on the tangential coordinate $\eta$.

Given the split \eqref{Ksplitd1}, the leading contribution to the determinant is $\det{(\d \delbar \KK)} = (\KK_1)_{\eta\etabar} (\KK_2)_{\lambda\lambdabar} + \OO(\epsilon^0)$, with $(\KK_1)_{\eta\etabar} (\KK_2)_{\lambda\lambdabar} = \OO(\epsilon^{-1})$.  Due to this, \eqref{sourcedMAd2} becomes a linear PDE for $\KK_2$ at leading order:
\begin{equation}\label{Poissond1}
\left[ \d_{y}^2 + 8 (\KK_1)_{\eta\etabar} \ \d_\lambda \delbar_{\lambdabar} \right] \KK_2 = - \frac{\tilde{Q}}{2\pi} \delta(y) \log{|\lambda|^2}~.
\end{equation}
$(\KK_1)_{\eta\etabar}$ is a function of $(\eta,\etabar)$ only and thus a constant from the point of view of the PDE.  The equation is solved by \eqref{K2gensol}.  Using this one determines the leading behaviour of the warp factor, \eqref{warpnb}.  We see that having a non-trivial induced metric on the brane worldvolume leads to a warp factor that does depend on the tangential coordinates.

\end{document}